\documentclass[%
  prx,
  twocolumn,
  superscriptaddress,
  amsmath,
  amssymb,
  aps
]{revtex4-2}
\usepackage[colorlinks = True, linkcolor = blue, citecolor = blue]{hyperref}
\usepackage{graphicx}
\usepackage[normalem]{ulem}
\usepackage{tikz}
\usetikzlibrary{arrows.meta, positioning}
\begin{document}

\title{Non-reciprocally interacting Ornstein-Uhlenbeck processes: Exceptional points, Anomalous relaxation, Pseudo-equilibrium and Boundary refrigeration }

\author{Soumya Kanti Pal}
\email{soumya.pal@tifr.res.in}
\affiliation{Department of Theoretical Physics, Tata Institute of Fundamental Research, Homi Bhabha Road, Mumbai 400005, India}

\author{Shamik Gupta}
\email{ shamik.gupta@theory.tifr.res.in}
\affiliation{Department of Theoretical Physics, Tata Institute of Fundamental Research, Homi Bhabha Road, Mumbai 400005, India}
\date{\today}

\begin{abstract}
Non-reciprocal interactions are ubiquitous in active, biological, and disordered systems, generically driving them out of equilibrium. Here, we introduce a hierarchy of non-reciprocally interacting Ornstein-Uhlenbeck (NROU) models governed by a tunable non-reciprocity parameter $g$. At a special point $g=g^*$, the drift matrix becomes non-diagonalizable, realizing exceptional points (EP's) of different orders, where eigenvalues and eigenvectors simultaneously coalesce. The hierarchy encompasses non-reciprocally coupled dimers, their disordered counterparts, and a many-body chain exactly mapping onto the paradigmatic Hatano-Nelson model in the arena of non-Hermitian quantum systems. For the disordered model, we show that the distribution of the EP location $g^*$ across disorder realizations develops a universal edge singularity precisely at the clean-system EP, and is manifestly non-self-averaging. Across all models, we find that at the EP, the usual exponential relaxation of the autocorrelation and covariance functions is dressed by a polynomial-in-time prefactor whose degree is set by the order of the EP and whose detailed structure encodes the spatial architecture of the chain. At complete asymmetry, the many-body chain exhibits ``pseudo-equilibrium'': its steady-state distribution factorizes into equilibrium-like single-particle measures despite a nonzero steady-state current. Moreover, the $N$-particle interacting system decomposes into $N/2$ independent complex OU processes. Finally, using the Harada-Sasa relation, we obtain a closed-form expression for the total steady-state heat dissipation and uncover a boundary refrigeration effect, in which the boundary particles switch from acting as a hot to a cold reservoir as the non-reciprocity is tuned. Together, these results show how simple, exactly solvable non-reciprocal models can pave the way for connecting non-Hermitian spectral theory and nonequilibrium stochastic thermodynamics.
\end{abstract}

\maketitle 
\tableofcontents
\section{Introduction}

Reciprocal interactions, in which the influence exerted by one degree of freedom on another is accompanied by a corresponding reciprocal response, form the foundation of equilibrium statistical mechanics and a large class of conventional many-body theories. Reciprocity is directly related to the existence of action-reaction symmetry encoding Newton's third law of motion. A wide range of physical systems, however, operate outside this framework, where the effective influence of one constituent on another need not be reciprocated. Such non-reciprocal interactions provide a generic mechanism for breaking detailed balance and generating collective dynamics far from equilibrium. These interactions arise naturally in driven and active matter~\cite{Soto2014,Ivlev2015,Scheibner2020,Fruchart2021,Gompper2025}, chemical and biological systems~\cite{Saha2019,Meredith2020,AgudoCanalejo2019,Saha2020,Hickey2023}, ecological~\cite{Carletti2022,Blumenthal2024,Ros2023,Weis2025} and neural networks~\cite{Poggialini2026,Badalotti2026}, social dynamics~\cite{Wieland2013}, engineered mechanical and metamaterial systems~\cite{Coulais2017,Librandi2021,Reisenbauer2024}, and many others. For a recent review, see Ref.~\cite{Fruchart2026}.

The consequences of reciprocity breaking extend across a remarkably diverse range of physical systems. In active and biological matter, non-reciprocal interactions can produce traveling waves~\cite{You2020}, oscillatory instabilities~\cite{Saha2020}, collective motion~\cite{Loos2023}, phase separation~\cite{Dinelli2023,Dopierala2025,Popli2025,Bandini2025}, and other forms of dynamical self-organization~\cite{You2020,Fruchart2021}. In active solids and metamaterials~\cite{Chen2021}, they give rise to odd elasticity~\cite{Zhou2026} and non-Hermitian mechanical responses~\cite{Scheibner2020PRL}, including unconventional elastic waves and spectral singularities~\cite{Scheibner2020}. Non-reciprocity has also emerged as an important ingredient in frustrated and disordered systems, where asymmetric interactions can generate spin-glass-like states, slow dynamics, aging, and oscillatory amorphous phases, with qualitatively distinct dynamical regimes~\cite{Hanai2024,GarciaLorenzana2025,GarciaLorenzana2025PRE}, complementing earlier studies of spin glasses with asymmetric interactions and pointing to a broader connection between non-reciprocity, frustration, and glassy physics~\cite{Drossel2000,Ros2023,Fournier2026}. Taken together, these results establish non-reciprocity as a generic mechanism capable of reorganizing relaxation, collective dynamics, and phase structure across systems ranging from biological and active matter to disordered many-body systems~\cite{Fruchart2021,Hanai2024}. A common feature underlying many of these phenomena is that their linearized dynamics is governed by a non-Hermitian generator, whose spectrum can exhibit exceptional points (EP's) at which eigenvalues and eigenvectors simultaneously coalesce. 

Exceptional points have been a well-established feature of non-Hermitian quantum systems~\cite{Heiss2012}, reshaping modern understanding of condensed matter phenomena~\cite {Kunst2018,Kawabata2019,Ashida2020,Bergholtz2021}. While there exist well-defined design principles~\cite{Pal2026} for constructing non-Hermitian quantum systems with tunable EP's of different orders, characterized by the number of coalescing eigenvectors, the corresponding scenario for classical systems is largely lacking, with some exceptions in the mathematical literature on defective Fokker--Planck operators~\cite{Fuhrman1995,Neerven2005,Arnold2014,Achleitner2015,Arnold2018}.   

Against this backdrop, it is natural to ask whether simple, analytically-tractable classical models can be constructed in which non-reciprocity naturally gives rise to EP's, and to investigate what consequences these singularities have for steady-state properties, both static and dynamic. Yet, despite growing interest in non-reciprocal stochastic systems, EP's of different orders have not, to the best of our knowledge, been systematically explored. We fill this gap by constructing a hierarchy of non-reciprocally interacting Ornstein--Uhlenbeck (NROU) models that realize EP's of different orders and host a broad range of striking phenomena, both during relaxation toward the steady state and in the steady state itself. The principal findings of this work are summarized below.

\subsection{Overview of the results}
Here, we lay out a summary of the main results along with the organization of the paper.

\begin{itemize}
    \item \textit{A hierarchy of non-reciprocally interacting Ornstein--Uhlenbeck processes hosting exceptional points}--- Ornstein--Uhlenbeck (OU) processes are foundational stochastic models for complex systems, with applications spanning microscopic, mesoscopic, and macroscopic scales. However, most studies of OU processes have focused on equilibrium dynamics, reciprocally-coupled driven dynamics, or dynamics with memory. Here, we introduce a hierarchy of non-reciprocally interacting OU processes governed by a single non-reciprocity parameter $g$. We show that at a special point $g=g^*$, the drift matrix underlying the OU dynamics becomes non-diagonalizable and assumes a Jordan canonical form, thereby giving rise to EP's. We start with a brief discussion on Jordan structures and how they lead to EP's of different orders (Sec.~\ref{Sec:Jordan-blocks}). 
    
    In our hierarchy, we consider three models: (i)~\underline{Model~I}, consisting of non-interacting dimers whose constituents are confined in harmonic potentials of identical stiffness constants and are coupled non-reciprocally; (ii)~\underline{Model~II}, consisting of non-interacting dimers whose constituents are confined in harmonic potentials with stiffness constants that are quenched-disordered random variables and are coupled non-reciprocally; and (iii)~\underline{Model~III}, a many-body generalization of Model~I in which particles are confined by harmonic potentials of identical stiffness, and which interact non-reciprocally with their nearest neighbours in one dimension with open boundaries. Model~III is in one-to-one correspondence with the paradigmatic Hatano--Nelson model of non-Hermitian quantum systems~\cite{Hatano1996}. Since non-reciprocity generically drives the system out of equilibrium, we further identify the parameter regimes in which a nonequilibrium steady state (NESS) exists for each of the three models (Sec.~\ref{Sec:Models}).

    \item \textit{Universality in the distribution of the EP location $g^*$}---
    Since the stiffness constants in Model~II are quenched-disordered, the location $g^*$ at which an individual disorder realization hits an EP fluctuates across realizations. We demonstrate that, irrespective of the underlying disorder distribution, the resulting distribution of $g^*$ develops a universal edge singularity precisely at the EP location of the corresponding clean (disorder-free) system, and the distribution is manifestly non-self-averaging (Sec.~ \ref{Sec:Edge-singularity}).

    \item \textit{Tunable anomalous relaxation at exceptional points: Temporal profile encoding spatial profile}---
    Across all models, we establish that at an EP, the usual exponential relaxation of the autocorrelation and covariance functions is dressed by a polynomial-in-time prefactor, whose degree is set by the order of the EP. In this work, we report  two major findings of such anomalous relaxation, e.g., for Model~III: 
    
    $(a)$ The relative autocorrelation function $\mathcal{R}_{\text{Auto}}(g,t)$ (defined in Eq.~\eqref{eq:relative-autocor-defn}) has the following temporal behaviour: 
    \begin{align}
        \mathcal{R}_{\text{Auto}} (g,t) \sim\begin{cases}
        &  \text{oscillatory};~g<g^*, \\
        &  \text{ $t^{\alpha}$ with $0\le \alpha\le N-1$ };~g=g^*,\\
        &  \text{exponential growth};~g>g^*, 
    \end{cases}
    \end{align}
    where $N$ is the system size. From the foregoing, we conclude that the relaxation dynamics at EP is highly intriguing, and, moreover, the largest value of the anomalous exponent $\alpha$ directly depends on $N$. Additionally, the value of $\alpha$ for a fixed $N$ depends crucially on the choice of the initial condition. 
    
    $(b)$ In the same vein, the relaxation dynamics of covariance or equal-time cross-correlation between two particles also gets polynomially dressed at EP. Specifically, we find that the cross-correlation between the $i$'th and $j$'th particles, see Eq.~\eqref{eq:covariance-matrix}, takes the form
    \begin{align}
    C_{ij}(g=g^*,t)& \sim \sum_{l=1}^{\min(i,j)} N_l \int_{0}^t \mathrm{d}t' ~ e^{-2kt'} (t')^{i+j-2l},
    \end{align} implying that the polynomial degree depends on the spatial separation, and, moreover, the structure of the polynomial encodes the spatial profile of the particle location for which the cross-correlation is being measured.
    Qualitatively similar results hold for other Models (Sec.~\ref{Sec:Anomalous relaxation}).

    \item \textit{Pseudo-equilibrium at complete asymmetry.}---
    At the special value of the non-reciprocity parameter $g=-1$, the drift matrix for Model~III possesses chiral symmetry.  We find that the steady-state distribution of the system factorizes into independent measures for individual particles that have the equilibrium form, leading to zero equal-time cross-correlation between the particles. Remarkably, the dynamics generates a nonzero, purely tangential steady-state probability current, furnishing an explicit example of a steady state that has the equilibrium Gibbs form and yet is a genuine NESS. Moreover, we show that at this special point, the $N$-particle interacting NROU dynamics can be exactly mapped to $N/2$ number of non-interacting complex OU processes (Sec.~\ref{Sec:Pseudo-Equilibrium}).   

    \item \textit{Heat dissipation: Boundary refrigeration}---
Using the Harada--Sasa relation~\cite{Harada2005}, which connects heat dissipation to the violation of the fluctuation--dissipation relation in an NESS, we compute the heat dissipated into the bath in the steady state. For Model~III, we obtain a closed-form expression for the total dissipated heat $\langle Q_{\text{tot}}\rangle$ in the thermodynamic limit. As expected, $\langle Q_{\text{tot}}\rangle$ remains positive, consistent with the second law of thermodynamics. At the level of individual particles, however, we uncover three distinct phenomena: $(a)$ at $g=-1$, the particles separate into two groups, namely, the bulk and the boundaries, with particles within each group exhibiting equal heat dissipation; $(b)$ at EP, $g=g^*$, the ratio of the heat dissipated by individual particles to that of the rightmost boundary particle~\footnote{In our convention, we denote $i=1$ lattice site to be the leftmost boundary, and $i=N$ site to be the rightmost boundary.} exhibits a divergence; and $(c)$ despite the total heat dissipated remaining positive, there exist parameter regimes in which one boundary particle acts as a source of heat, dissipating energy into the bath and thereby exhibiting positive heat dissipation, while the other acts as a sink, absorbing heat from the bath and thereby exhibiting negative heat dissipation. Interestingly, the sign of the boundary heat dissipation switches around the equilibrium point $g=1$. Thus, while the bulk particles act as effective hot reservoirs for all values of the non-reciprocity parameter, the boundary can undergo a transition from a hot to a cold reservoir, thereby realizing boundary refrigeration via tuning of $g$ (Sec.~\ref{Sec:Heat-dissipation}).         
\end{itemize}


\section{Basics of exceptional points and Jordan canonical forms} \label{Sec:Jordan-blocks}

We first review the basics of exceptional points (EP's), which have garnered much attention in the quantum domain pertaining to the studies of non-Hermitian Hamiltonians modeling the physics of open quantum systems~\cite{Ashida2020}. For any general $(N\times N)$ matrix $A$ to host EP, it must have simultaneous coalescence of eigenvalues and eigenvectors. To put it concretely, if $A$ has an $m$-fold degenerate eigenvalue $\lambda$, with the corresponding eigenvector equation
\begin{align}
    (A - \lambda \mathbb{I}_N  )|v \rangle =0 
\end{align}
yielding $d<m$ number of independent eigenvectors $|v\rangle $, then $A$ is said to host an EP of order $(m-d+1)$ corresponding to $\lambda$, with $(m-d+1)$ giving the number of coalescing eigenvectors. In the literature, $m$ is called the algebraic multiplicity of the eigenvalue $\lambda$, whereas $d$ is known as the geometric multiplicity, and therefore, the EP condition $d<m$ implies that the geometric multiplicity is strictly less than the algebraic multiplicity. As a result, such matrices are non-diagonalizable.
Moreover, the existence of EP is strictly conditioned on the non-Hermiticity of $A$, i.e., $A=A^\dagger$ (or $A\neq A^T$ if $A$ is real). For Hermitian matrices, one can prove that $d=m$, implying non-existence of EP; for details, see Appendix~\ref{APP:EP-absence-Hermitian} \footnote{Note that non-Hermiticity of a matrix is necessary but not sufficient for hosting EP's.}.

We now delve into the underlying structure of matrices that host EP's. Such matrices are related by a similarity transformation to Jordan canonical forms that characterize EP's of different orders~\cite{Horn1985}. It is evident that for an $(N\times N)$ matrix, the order of EP is upper-bounded by $N$. We now show how one can systematically construct all possible Jordan canonical forms corresponding to all possible EP's in arbitrary dimensions $N$. 

For $N=2$, the EP can be only of order $2$ with the corresponding Jordan block $J_2(\lambda) = \begin{pmatrix}
    \lambda & 1 \\
    0 & \lambda 
\end{pmatrix}$. Next, for $N=3$, the possible order of EP's can be either $3$ or $2$, and the corresponding Jordan Blocks are 
\begin{equation}
    J_3(\lambda) = \begin{pmatrix}
    \lambda & 1 & 0 \\
    0 & \lambda & 1 \\
    0 & 0 & \lambda
\end{pmatrix}, ~J_2(\lambda_1) \oplus J_1(\lambda_2) = \begin{pmatrix}
    \lambda_1 & 1 & 0 \\
    0 & \lambda_1 & 0 \\
    0 & 0 & \lambda_2
\end{pmatrix},
\end{equation}
where $J_1(\lambda)$ is simply a number corresponding to $N=1$. 
Subsequently, for $N=4$, the possible Jordan blocks are 
\begin{align}
    & J_4(\lambda) = \begin{pmatrix}
        \lambda & 1 & 0 & 0 \\
    0 & \lambda & 1 & 0 \\
    0 & 0 & \lambda & 1\\
    0 & 0 & 0 & \lambda
    \end{pmatrix}; ~\text{EP of order $4$},\\ 
    & J_3(\lambda_1) \oplus J_1(\lambda_2) = \begin{pmatrix}
        \lambda_1 & 1 & 0 & 0 \\
    0 & \lambda_1 & 1 & 0 \\
    0 & 0 & \lambda_1 & 0\\
    0 & 0 & 0 & \lambda_2
    \end{pmatrix};~ \text{EP of order $3$}, \\
    & J_2(\lambda_1) \oplus J_2(\lambda_2) = \begin{pmatrix}
        \lambda_1 & 1 & 0 & 0 \\
    0 & \lambda_1 & 0 & 0 \\
    0 & 0 & \lambda_2 & 1\\
    0 & 0 & 0 & \lambda_2 
    \end{pmatrix};~ \text{two EP's of order $2$}, \\
    & J_2(\lambda_1) \oplus J_1(\lambda_2) \oplus J_1(\lambda_3) \nonumber \\ 
    &  = \begin{pmatrix}
        \lambda_1 & 1 & 0 & 0 \\
    0 & \lambda_1 & 0 & 0 \\
    0 & 0 & \lambda_2 & 0\\
    0 & 0 & 0 & \lambda_3 
    \end{pmatrix};~\text{EP of order $2$}.
\end{align}

More generally, the Jordan canonical form of an $N\times N$ non-Hermitian matrix is classified by partitions $\{i_1,\dots,i_k\}$ of $N$,
\begin{align}
\bigoplus_{j=1}^{k}J_{i_j}(\lambda_j),
\qquad
\sum_{j=1}^{k}i_j=N .
\end{align}
The order of an EP is determined by the size of the largest Jordan block,
\begin{align}
O_{\rm EP}=\max(i_j).
\end{align}
Therefore, an EP occurs whenever at least one block satisfies $i_j\geq2$, whereas the trivial partition $\{1,1,\dots,1\}$ corresponds to a diagonalizable matrix.

The highest-order EP, which is of order $N$, is represented by the Jordan block
\begin{align}
J_N(\lambda)=
\begin{pmatrix}
\lambda&1&0&\cdots&0\\
0&\lambda&1&\cdots&0\\
0&0&\lambda&\ddots&0\\
\vdots&\vdots&\ddots&\ddots&1\\
0&0&0&\cdots&\lambda
\end{pmatrix}.
\end{align}
We may write $J_N(\lambda)=\lambda\mathbb{I}+J_N(0)$, where $J_N(0)$ is a nilpotent matrix satisfying
\begin{align}
[J_N(0)]^N=0,
\qquad
[J_N(0)]^{N-1}\neq0 .
\end{align}
This nilpotency relation will be crucially exploited in the later part of the paper. 

\section{Non-reciprocal Models} \label{Sec:Models} 

In this section, we introduce a hierarchy of non-reciprocally interacting Ornstein–Uhlenbeck (NROU) processes that host EP's, with progressively richer models retaining the same underlying physics while revealing additional dynamical features. We elucidate how this hierarchy is connected to the Jordan-block structure discussed in the previous section. We then analyze the conditions for the existence of a nonequilibrium steady state (NESS) in these systems and show that exceptional points arise naturally upon tuning the non-reciprocity parameter. To set the stage, we briefly review the multivariate Ornstein–Uhlenbeck (OU) process.

\subsection{A brief review of the multivariate OU process}
The standard Ornstein--Uhlenbeck (OU) process describes overdamped Langevin dynamics of $N$ interacting particles in one dimension, with each particle confined in a harmonic potential and the entire system immersed in a heat bath in equilibrium. Denoting by $x_i(t)$ the position of the $i$'th particle with respect to the center of the respective harmonic potential, the dynamics written compactly in terms of the $(N\times 1)$ state vector $\vec{x}(t)=(x_1(t),x_2(t),\ldots,x_N(t))^T$ reads as~\cite{vanKampen1992}
\begin{align} \label{eq:multi-variate OU}
    \frac{d\vec{x}(t)}{dt}=M\vec{x}(t)+\vec{\eta}(t),
\end{align}
where $M$ is a constant drift matrix encoding both the interactions between particles and the harmonic trapping potential for individual particles, while $\vec{\eta}(t)$ is a Gaussian white-noise term with strength $D$, characterized by
\begin{align}
    \langle \vec{\eta}(t)\vec{\eta}^{T}(t')\rangle
    =2D\,\mathbb{I}_N\,\delta(t-t').
\end{align}
The solution to Eq.~\eqref{eq:multi-variate OU} is 
\begin{align} \label{eq:coordiante soultion}
    \vec{x}(t) = e^{Mt} \vec{x}(0) + \int_{0}^t ds ~ \eta(s) e^{M (t-s)}.
\end{align}
The stability of this solution is directly related to the eigenvalue structure of the drift matrix $M$; in particular, stability demands that the real part of all eigenvalues of $M$ must be zero or negative, i.e., 
\begin{align}
    \Re[\lambda_i] \le 0 ,~ M |v_i\rangle = \lambda_i |v_i \rangle. 
\end{align}
The corresponding joint probability distribution $P(\vec{x},t)$ of the positions of the particles approaches a steady state in the asymptotic limit $t \to \infty$, given by the Gaussian distribution~\cite{vanKampen1992} 
\begin{align} \label{eq:Gaussian-stationary-solution}
    P_{\mathrm{ss}}(\vec{x})
    =
    \frac{1}{(2\pi)^{N/2}\sqrt{\det C}}
    \exp\left[
    -\frac{1}{2}\vec{x}^{\,T}C^{-1}\vec{x}
    \right],
\end{align}
where $C$ is the symmetric steady-state covariance matrix with its elements $C_{ij}=\langle x_i x_j \rangle_{\mathrm{ss}}$, and the associated probability current in the state space is given by 
\begin{align} \label{eq:steady-state-current}
    \vec{J}_{\mathrm{ss}}(\vec{x})= M\vec{x}P_{\mathrm{ss}}(\vec{x}) - D \vec\nabla P_{\mathrm{ss}}(\vec{x}).
\end{align}
The current in Eq.~\eqref{eq:steady-state-current} determines the nature of the steady state as to whether the system is in equilibrium or in an NESS. 

In equilibrium, the steady-state current vanishes, i.e., $\vec{J}_{\mathrm{ss}}(\vec{x}) =0$. This is a consequence of the principle of detailed balance that holds in equilibrium, which in turn conditions the drift matrix $M$ to be symmetric, i.e., $M= M^T$, which follows since $M= -D C^{-1}$, upon substitution of Eq.~\eqref{eq:Gaussian-stationary-solution} into Eq.~\eqref{eq:steady-state-current}. The corresponding equilibrium measure is proportional to $ \exp{[-V(\vec{x})/D]}$, where the potential is $V(\vec x) = \frac{1}{2} \vec x^T M \vec x$. For a single particle, the potential reduces to $V(x) = (1/2) k x^2$, where $k>0$ is the stiffness. Moving to the multivariate case, we have $\{M_{ii} =-k_i\}_{ 1 \le i \le N}$, with $k_i$'s being the stiffness of the harmonic potential for individual particles, and $M_{ij} = M_{ji}$ represents the interaction between the $i$'th and the $j$'th particle. 

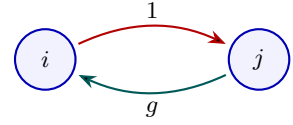
\begin{figure}[tbp]
\centering
\begin{tikzpicture}[
    mode/.style={
        circle,
        draw=blue!70!black,
        fill=blue!5,
        thick,
        minimum size=0.8cm,
        font=\small\bfseries
    },
    interaction/.style={
        -{Stealth[scale=1.0]},
        thick,
        shorten >=2pt,
        shorten <=2pt
    }
]

    \node[mode] (i) {$i$};
    \node[mode] (j) [right=2cm of i] {$j$};

    \path[interaction, bend left=25, draw=red!70!black] (i) edge node[above, font=\small] {$1$} (j);

    \path[interaction, bend left=25, draw=teal!70!black] (j) edge node[below, font=\small] {$g$} (i);

\end{tikzpicture}
\caption{Schematic representation of the non-reciprocal coupling between the particles $i$ and $j$, where $g$ is the non-reciprocity strength. }
\label{fig:non-reciprocal_ij}
\end{figure}

While the steady-state distribution in Eq.~\eqref{eq:Gaussian-stationary-solution} holds for any choice of $M$, arguably, the most interesting case is 
\begin{align}
    M_{ij} \neq M_{ji}, ~ M_{ij}\in \mathbb{R},
\end{align}
leading us directly to the arena of non-reciprocal systems. From the above discussion, it is evident that, if and when a steady state exists, it ought to be an NESS with $\vec{J}_{\mathrm{ss}} \neq 0 $. At the microscopic level, the asymmetry of $M$ implies a breakdown of reciprocity between particle interactions: particle $i$ influences particle $j$ differently from how particle $j$ influences particle $i$. For instance, particle $i$ may interact with particle $j$ with coupling strength $1$, whereas the reverse interaction occurs with coupling strength $g$, as illustrated in Fig.~\ref{fig:non-reciprocal_ij}. Henceforth, we will refer to $g$ as the strength of non-reciprocity. In this backdrop, we now discuss our hierarchy of non-reciprocally interacting OU models. We subsequently uncover how such non-reciprocal couplings leads to novel NESS and exotic dynamical phenomena.

\subsection{Model I: Non-interacting dimers in uniform trapping potential}
We consider the total number of particles $N$ to be even, and group them into $N/2$ number of pairs (or dimers) as 
\begin{align}
    \{(1,2), (3,4), (5,6), \ldots, (N-1,N)\}.
\end{align}
 Clearly, such grouping is permutationally invariant, and therefore, there is no a priori reason to consider the pairs in $(i,i+1)$ form. However, such a consideration becomes meaningful when the index $i$ is interpreted as the site index of a one-dimensional lattice.

\begin{figure}[t]
\centering
\begin{tikzpicture}[
    mode/.style={
        circle,
        draw=blue!70!black,
        fill=blue!5,
        thick,
        minimum size=0.65cm,
        font=\scriptsize\bfseries,
        inner sep=1pt
    },
    interaction/.style={
        -{Stealth[scale=0.8]},
        thick,
        shorten >=1.5pt,
        shorten <=1.5pt
    }
]

    
    \draw[domain=-0.2:1.1, smooth, variable=\x, draw=red!50, very thick] 
        plot ({\x}, {1.3 - 1.2*cos(220*(\x - 0.5))});
        
    \draw[domain=1.0:2.1, smooth, variable=\x, draw=red!50, very thick] 
        plot ({\x}, {1.3 - 1.2*cos(220*(\x - 1.6))});
        
    \draw[domain=2.0:3.1, smooth, variable=\x, draw=red!50, very thick] 
        plot ({\x}, {1.3 - 1.2*cos(220*(\x - 2.55))});
        
    \draw[domain=3.1:4.2, smooth, variable=\x, draw=red!50, very thick] 
        plot ({\x}, {1.3 - 1.2*cos(220*(\x - 3.65))});
        
    \draw[domain=4.5:5.6, smooth, variable=\x, draw=red!50, very thick] 
        plot ({\x}, {1.3 - 1.2*cos(220*(\x - 5.05))});
        
    \draw[domain=5.5:6.7, smooth, variable=\x, draw=red!50, very thick] 
        plot ({\x}, {1.3 - 1.2*cos(220*(\x - 6.15))});

    
    \node[mode] (m1) at (0.5, 1.1) {$1$};
    \node[mode] (m2) at (1.6, 1.1) {$2$};
    \path[interaction, bend left=24, draw=red!70!black] (m1) edge node[above, font=\small] {$1$} (m2);
    \path[interaction, bend left=24, draw=teal!70!black] (m2) edge node[below, font=\small] {$g$} (m1);
    
    \node[mode] (m3) at (2.55, 1.1) {$3$};
    \node[mode] (m4) at (3.65, 1.1) {$4$};
    \path[interaction, bend left=24, draw=red!70!black] (m3) edge node[above, font=\small] {$1$} (m4);
    \path[interaction, bend left=24, draw=teal!70!black] (m4) edge node[below, font=\small] {$g$} (m3);
    
    \node[text=black!40, font=\bfseries] at (4.35, 1.1) {$\cdots$};

    \node[mode] (mN1) at (5.05, 1.1) {\!N-1\!};
    \node[mode] (mN)  at (6.15, 1.1) {$N$};
    \path[interaction, bend left=24, draw=red!70!black] (mN1) edge node[above, font=\small] {$1$} (mN);
    \path[interaction, bend left=24, draw=teal!70!black] (mN)  edge node[below, font=\small] {$g$} (mN1);
    
\end{tikzpicture}
\caption{Schematic representation of Model~I: Non-interacting dimers with non-reciprocal interactions between their constituents that are confined in a harmonic potential of identical stiffness.}
\label{fig:schematic-model I}
\end{figure}
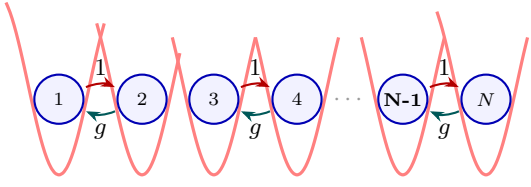

Now, for the first model in the hierarchy, we consider the drift matrix in Eq.~\eqref{eq:multi-variate OU} to have the following form: 
\begin{equation}\label{eq:Model I-drift-matrix}
M(g,k)= 
\begin{pmatrix}
-k & 1 & 0 & 0 & \cdots & 0 & 0\\
g & -k & 0 & 0 & \cdots & 0 & 0\\
0 & 0 & -k & 1 & \cdots & 0 & 0\\
0 & 0 & g & -k & \cdots & 0 & 0\\
\vdots & \vdots & \vdots & \vdots & \ddots & 1 & 0\\
0 & 0 & 0 & 0 & \cdots & -k & 1\\
0 & 0 & 0 & 0 & \cdots & g & -k
\end{pmatrix}_{N\times N},
\end{equation}
where $g$ is the non-reciprocal interaction strength between the $i$'th and $(i+1)$'th  particles forming a pair, for odd $i$. For $g<0$, the interaction from particle $i$ to particle $(i+1)$ is attractive, whereas the reverse interaction is repulsive. The drift matrix~\eqref{eq:Model I-drift-matrix}, together with Eq.~\eqref{eq:multi-variate OU}, describes the noisy dynamics of $N/2$ dimers, with each particle confined in a harmonic potential of identical stiffness $k>0$ and no intra-dimer interaction, as illustrated in Fig.~\ref{fig:schematic-model I}.
This latter feature reduces the system to one involving $N/2$ non-interacting dimers, with
\begin{align}\label{Eq:Block-decomposition-model I}
    M (g,k) = \bigoplus_{i \in \mathrm{Odd}} M^{(i,i+1)}, ~ M^{(i,i+1)}= \begin{pmatrix}
        -k & 1 \\
        g & -k 
    \end{pmatrix}.
\end{align}
Thus, the problem, in effect, becomes that of a single dimer described by the matrix $M^{(i,i+1)}$.

Now, the existence of NESS demands the real part of the eigenvalues of the matrix $M^{(i,i+1)}$, namely, $\lambda_{\pm} = -k \pm \sqrt{g}$, to be zero or negative, leading to the condition that $g \le k^2$. Following Eq.~\eqref{Eq:Block-decomposition-model I}, the corresponding steady-state distribution~\eqref{eq:Gaussian-stationary-solution} for the entire system will factorize into stationary Gaussian distributions for individual dimers.
The EP of order $2$ for individual dimer systems occurs when the non-reciprocity strength $g$ is tuned to $g^*=0$, resulting in the Jordan block structure $M^{(i,i+1)}(g=g^*)=-k \mathbb{I}_2 + J_2^{(i,i+1)}(0) $. Subsequently, at $g=g^*$, the entire system admits the following Jordan block decomposition  
\begin{align}
    M(g=g^*,k) = \bigoplus_{i \in \mathrm{Odd}} \left[-k \mathbb{I}_2 + J_2^{(i,i+1)}(0)\right], 
\end{align}
implying the occurrence $N/2$ number of EP's of order 2.

\subsection{Model II:  Non-interacting dimers in disordered trapping potential}
\begin{figure}[t]
\centering
\begin{tikzpicture}[
    mode/.style={
        circle,
        draw=blue!70!black,
        fill=blue!5,
        thick,
        minimum size=0.65cm,
        font=\scriptsize\bfseries,
        inner sep=1pt
    },
    interaction/.style={
        -{Stealth[scale=0.8]},
        thick,
        shorten >=1.5pt,
        shorten <=1.5pt
    }
]

    \draw[domain=-0.2:1.1, smooth, variable=\x, draw=red!50, very thick] 
        plot ({\x}, {1.3 - 1.1*cos(200*(\x - 0.5))});
        
    \draw[domain=1.0:2.1, smooth, variable=\x, draw=red!50, very thick] 
        plot ({\x}, {1.3 - 0.5*cos(220*(\x - 1.6))});
        
    \draw[domain=2.0:3.1, smooth, variable=\x, draw=red!50, very thick] 
        plot ({\x}, {1.3 - 0.8*cos(220*(\x - 2.55))});
        
    \draw[domain=3.1:4.2, smooth, variable=\x, draw=red!50, very thick] 
        plot ({\x}, {1.3 - 1.2*cos(220*(\x - 3.65))});
        
    \draw[domain=4.5:5.6, smooth, variable=\x, draw=red!50, very thick] 
        plot ({\x}, {1.3 - 0.4*cos(220*(\x - 5.05))});
        
    \draw[domain=5.5:6.7, smooth, variable=\x, draw=red!50, very thick] 
        plot ({\x}, {1.3 - 0.9*cos(200*(\x - 6.15))});

    
    \node[mode] (m1) at (0.5, 1.6) {$1$};
    \node[mode] (m2) at (1.6, 1.6) {$2$};
    \path[interaction, bend left=24, draw=red!70!black] (m1) edge node[above, font=\small] {$1$} (m2);
    \path[interaction, bend left=24, draw=teal!70!black] (m2) edge node[below, font=\small] {$g$} (m1);
    
    \node[mode] (m3) at (2.55, 1.6) {$3$};
    \node[mode] (m4) at (3.65, 1.6) {$4$};
    \path[interaction, bend left=24, draw=red!70!black] (m3) edge node[above, font=\small] {$1$} (m4);
    \path[interaction, bend left=24, draw=teal!70!black] (m4) edge node[below, font=\small] {$g$} (m3);
    
    \node[text=black!40, font=\bfseries] at (4.35, 1.6) {$\cdots$};

    \node[mode] (mN1) at (5.05, 1.6) {\!N-1\!};
    \node[mode] (mN)  at (6.15, 1.6) {$N$};
    \path[interaction, bend left=24, draw=red!70!black] (mN1) edge node[above, font=\small] {$1$} (mN);
    \path[interaction, bend left=24, draw=teal!70!black] (mN)  edge node[below, font=\small] {$g$} (mN1);
    
\end{tikzpicture}
\caption{Schematic representation of Model~II: Non-interacting dimers with non-reciprocal interactions between their constituents that are confined in harmonic potentials of random stiffness.}
\label{fig:schematic-model-II}
\end{figure}
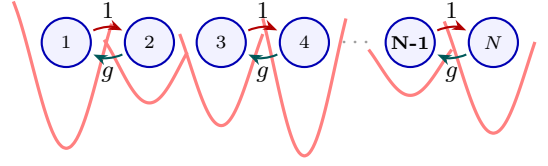

The next model in the hierarchy consists of individual particles confined in harmonic potentials with disordered stiffness constants $k_i>0$, which we take to be independent and identically-distributed random variables drawn from a common distribution $f(k)$. The positivity of $k_i$'s puts the following restriction on the support of $f(k)$:
\begin{align}
    f(k)  \begin{cases}
        =0  \text{ if } k \le 0, \\
        \neq 0  \text{ if } k >0. 
    \end{cases}
\end{align}
The interaction structure remains the same as in Model I, whereby the particles forming the dimers are interacting non-reciprocally, but there is no interaction between different dimers. A schematic of the same is illustrated in Fig.~\ref{fig:schematic-model-II}. The block decomposition in Eq.~\eqref{Eq:Block-decomposition-model I} of the drift matrix for the entire system also holds here, with the disordered stiffness now rendering the drift matrix for the individual dimers to be
\begin{align}\label{eq:Drfit-matrix-model-II}
    M^{(i,i+1)} = \begin{pmatrix}
        -k_i & 1 \\
        g & -k_{i+1} 
    \end{pmatrix},
\end{align}
with the eigenvalues $\lambda_{\pm}^{(i,i+1)} = \left[-(k_i + k_{i+1} ) \pm \sqrt{(k_i-k_{i+1})^2 + 4g}\right]/ 2$. The condition for the eigenvalues to have negative or zero real part for the existence of NESS requires that  $g \le \min_{i}\{k_ik_{i+1}\}$. Similar to Model I, for a given realization of $\{k_i\}$, the steady-state distribution~\eqref{eq:Gaussian-stationary-solution} here also factorizes into Gaussian measures for individual dimers.

Now, for a given realization $\{k_i\}$, the drift matrix $M^{(i,i+1)}$ hosts EP at $g^*= - (k_i - k_{i+1})^2/4$, whereby  the eigenvalues $\lambda_{\pm}^{(i,i+1)}$ become degenerate, and $M^{(i,i+1)}(g=g^*)$ becomes related to the Jordan canonical form via a similarity transformation $S$:
\begin{align}
    & M^{(i,i+1)} = S \left[ -(k_i + k_{i+1})/2~ \mathbb{I}_2 + J_2(0) \right] S^{-1}, \\
    & S= \begin{pmatrix}
        (k_{i+1} - k_i)/2 & 0 \\
        1 & 1 
    \end{pmatrix}.  
\end{align}
However, unlike Model I, the value of $g^*$ crucially depends on the individual dimer. Therefore, even though, the drift matrix $M$ has a block-diagonal structure, individual blocks hit EP at different values of the non-reciprocity parameter. This motivates us to look into the distribution $p(g^*)$ of EP location, which we discuss below.

\subsubsection{Edge singularity in the distribution of EP locations} \label{Sec:Edge-singularity}

\begin{figure}[tbp]
    \centering
    \includegraphics[width=1\linewidth]{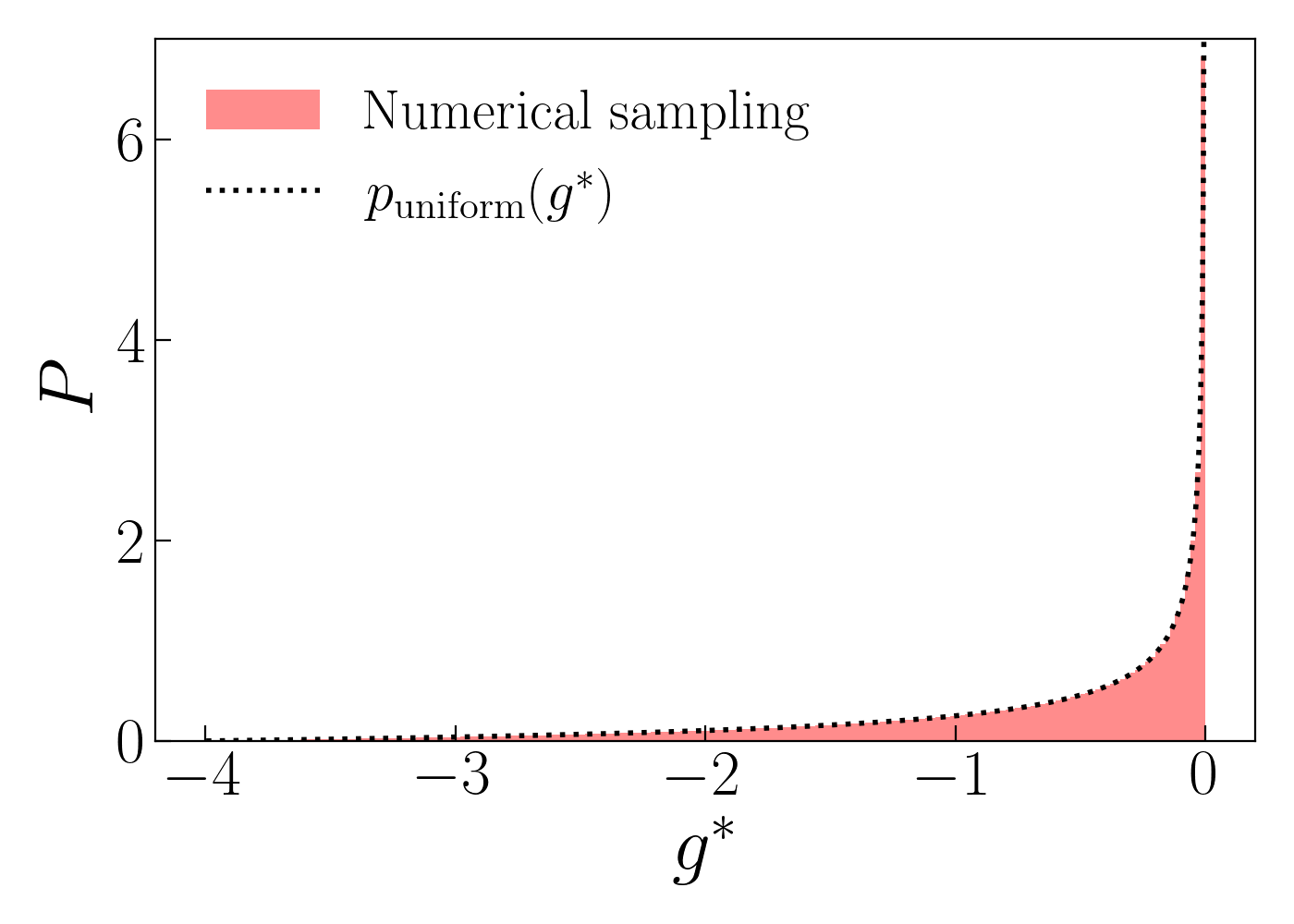}
    \caption{For Model~II, with the stiffness constant $k_i$'s drawn uniformly from the interval $[1,5]$, the figure shows the distribution of the EP location $g^*$ computed numerically for $10^8$ disorder realizations and compared against our theoretical prediction, Eq.~\eqref{eq:theory-pred}. 
}
    \label{fig:Edge-singularity}
\end{figure}

Using $g^* =- (k_i - k_{i+1})^2/4$ and  conservation of probability, we find the expression of $p(g^*)$ to be
\begin{align}
    p(g^*) = \Theta(-g^*) \frac{2}{\sqrt{-g^*}}  \int_{2 \sqrt{-g^*}}^\infty dk ~ f(k) f(k - 2\sqrt{-g^*}),
\end{align}
where $\Theta(x) = \begin{cases}
    1, \text{ if } x\ge 0\\
    0, \text{ if } x< 0 
\end{cases}$.
This immediately leads to the revelation that for any parent distribution of the stiffness $k_i$'s, the distribution $p(g^*)$ exhibits a remarkable edge singularity as $g^* \to 0^-$. Physically, this implies that the disordered system statistically favours the EP location of the clean system (Model~I) by creating a singularity at $g^*=0$. In other words, although the EP location of an individual disorder realization is shifted away from the clean-system value by the local fluctuations in the diagonal elements of the drift matrix, the ensemble of disorder realizations retains a pronounced statistical preference for the clean-system EP. Moreover, the distribution is non-self-averaging: the mean of the distribution and the most-probable value do not coincide. To demonstrate, we consider the representative choice of $f(k)$ to be a uniform distribution in the range $[a,b]$ with $b>a>0$. The corresponding distribution of the EP location is 
\begin{align}\label{eq:theory-pred}
    p_{\text{uniform}}(g^*) = 
    \frac{2}{y^2}
    \left(
        \frac{y}{\sqrt{-g^*}} - 2
    \right)
    \Theta(-g^*)
    \Theta\left(
        g^* + \frac{y^2}{4}
    \right),
\end{align}
with $y \equiv (a -b)$. Figure~\ref{fig:Edge-singularity} illustrates an excellent agreement between the numerical sampling and the theoretical prediction of the distribution of $g^*$, highlighting the edge singularity.

In passing, we remark that if the $k_i$'s were sampled from a discrete distribution such as a normalized sum of a finite number of delta functions, as opposed to a continuous distribution, there would be a finite probability for multiple dimer blocks to host EP at the same $g^*$. This underlines a crucial distinction between the nature of the disorder distribution, although the predicted edge singularity would survive in all cases.

\subsection{Model III: Interacting particles in uniform trapping potential and Hatano-Nelson model}

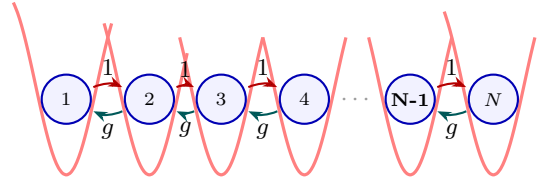
\begin{figure}[t]
\centering
\begin{tikzpicture}[
    mode/.style={
        circle,
        draw=blue!70!black,
        fill=blue!5,
        thick,
        minimum size=0.65cm,
        font=\scriptsize\bfseries,
        inner sep=1pt
    },
    interaction/.style={
        -{Stealth[scale=0.8]},
        thick,
        shorten >=1.5pt,
        shorten <=1.5pt
    }
]

    
    \draw[domain=-0.2:1.1, smooth, variable=\x, draw=red!50, very thick] 
        plot ({\x}, {1.3 - 1.2*cos(220*(\x - 0.5))});
        
    \draw[domain=1.0:2.1, smooth, variable=\x, draw=red!50, very thick] 
        plot ({\x}, {1.3 - 1.2*cos(220*(\x - 1.6))});
        
    \draw[domain=2.0:3.1, smooth, variable=\x, draw=red!50, very thick] 
        plot ({\x}, {1.3 - 1.2*cos(220*(\x - 2.55))});
        
    \draw[domain=3.1:4.2, smooth, variable=\x, draw=red!50, very thick] 
        plot ({\x}, {1.3 - 1.2*cos(220*(\x - 3.65))});
        
    \draw[domain=4.5:5.6, smooth, variable=\x, draw=red!50, very thick] 
        plot ({\x}, {1.3 - 1.2*cos(220*(\x - 5.05))});
        
    \draw[domain=5.5:6.7, smooth, variable=\x, draw=red!50, very thick] 
        plot ({\x}, {1.3 - 1.2*cos(220*(\x - 6.15))});

    
    \node[mode] (m1) at (0.5, 1.1) {$1$};
    \node[mode] (m2) at (1.6, 1.1) {$2$};
    \path[interaction, bend left=24, draw=red!70!black] (m1) edge node[above, font=\small] {$1$} (m2);
    \path[interaction, bend left=24, draw=teal!70!black] (m2) edge node[below, font=\small] {$g$} (m1);
    
    \node[mode] (m3) at (2.55, 1.1) {$3$};
    \path[interaction, bend left=24, draw=red!70!black] (m2) edge node[above, font=\small] {$1$} (m3);
    \path[interaction, bend left=24, draw=teal!70!black] (m3) edge node[below, font=\small] {$g$} (m2);

    \node[mode] (m4) at (3.65, 1.1) {$4$};
    \path[interaction, bend left=24, draw=red!70!black] (m3) edge node[above, font=\small] {$1$} (m4);
    \path[interaction, bend left=24, draw=teal!70!black] (m4) edge node[below, font=\small] {$g$} (m3);
    
    \node[text=black!40, font=\bfseries] at (4.35, 1.1) {$\cdots$};

    \node[mode] (mN1) at (5.05, 1.1) {\!N-1\!};
    \node[mode] (mN)  at (6.15, 1.1) {$N$};
    \path[interaction, bend left=24, draw=red!70!black] (mN1) edge node[above, font=\small] {$1$} (mN);
    \path[interaction, bend left=24, draw=teal!70!black] (mN)  edge node[below, font=\small] {$g$} (mN1);
    
\end{tikzpicture}
\caption{Schematic representation of Model~III: A one-dimensional open lattice of particles with non-reciprocal interactions, and individual particles confined by a harmonic potential of identical stiffness.}
\label{fig:schematic-model III}
\end{figure}

The most non-trivial model in the hierarchy is to consider every particle to be interacting non-reciprocally with both its immediate neighbours, with individual particles trapped in a harmonic potential of identical stiffness. Owing to the extended interaction structure, this model cannot be reduced to a simple dimerized structure as in Models~I, II. A schematic of the model is shown in Fig.~\ref{fig:schematic-model III}. The corresponding drift matrix is 
\begin{align} \label{eq:Model III-drift matrix}
    M(g,k)= \left( \begin{array}{cccccccccc}
-k & 1 & 0 & \cdots & 0 & 0 & 0 & \cdots & 0 & 0 \\
g & -k & 1 & \cdots & 0 & 0 & 0 & \cdots & 0 & 0 \\
0 & g & -k & \ddots & \vdots & \vdots & \vdots & \ddots & \vdots & \vdots \\
\vdots & \vdots & \ddots & -k & 1 & 0 & 0 & \cdots & 0 & 0 \\
0 & 0 & \cdots & g & -k & 1 & 0 & \cdots & 0 & 0 \\
0 & 0 & \cdots & 0 & g & -k & 1 & 0 & \cdots & 0 \\
0 & 0 & \cdots & 0 & 0 & g & -k & 1 & \cdots & 0 \\
\vdots & \vdots & \ddots & \vdots & \vdots & \vdots & \ddots & -k & \ddots & \vdots \\
0 & 0 & \cdots & 0 & 0 & 0 & \cdots & g & -k & 1 \\
0 & 0 & \cdots & 0 & 0 & 0 & \cdots & 0 & g & -k
\end{array} \right),
\end{align}
which has an embedded Jordan-Block structure at $g^*=0$: 
\begin{align} \label{eq:Jordan-decomposition-Model-III}
    M(g=g^*,k) = -k \mathbb{I}_N + J_N(0).
\end{align}
This immediately implies that at $g=g^*$, the drift matrix hosts an EP of order $N$.

The present setup can also be interpreted as a lattice model, and the matrix in Eq.~\eqref{eq:Model III-drift matrix} is the non-Hermitian Hamiltonian of a single-particle quantum system, namely, the well-known Hatano-Nelson model~\cite{Hatano1996}, given by
\begin{align} \label{eq:Hatano-Nelson}
    \hat{H}_{\text{HN}} = \sum_{j=1}^{N-1} \biggr[ |j\rangle \langle j+1| + g |j+1\rangle \langle j|   \biggr] - k\sum_{j=1}^N   |j\rangle \langle j| .
\end{align}
This model is a non-reciprocal generalization of the paradigmatic tight-binding Hamiltonian with a constant onsite potential; here the hopping probability of the particle  to move to the left is different from the one to the right. Note that the boundary condition has been taken to be open, and a detailed discussion on this specific choice will be made later.         

The NESS existence condition for this model is simply $\max_{i}\left( \Re[\lambda_i]\right)\le 0$, where $\{\lambda_i\}$ are the eigenvalues of $M$ in Eq.~\eqref{eq:Model III-drift matrix}. The spectrum of $M$ is in one-to-one correspondence with that of the Hatano--Nelson model in Eq.~\eqref{eq:Hatano-Nelson}, and is given by
\begin{align}
    \lambda_n = -k + 2\sqrt{g}\cos\left(\frac{n\pi}{N+1}\right),
    \qquad n=1,\ldots,N,
\end{align}
see Appendix~\ref{APP:Hatano-Nelson-Spectrum} for details. Therefore, as $N\to \infty$ limit, one must take $k^2 \ge 4g$ to ensure the existence of NESS. For any finite $N$, this condition is sufficient, as can be verified by setting $N=2$, for which the model reduces to Model~I. Evidently, the NESS measure of the model continues to be the Gaussian in Eq.~\eqref{eq:Gaussian-stationary-solution}.

\subsubsection{Sensitivity to boundary condition}
\begin{figure}[tbp]
    \centering
    \includegraphics[width=1\linewidth]{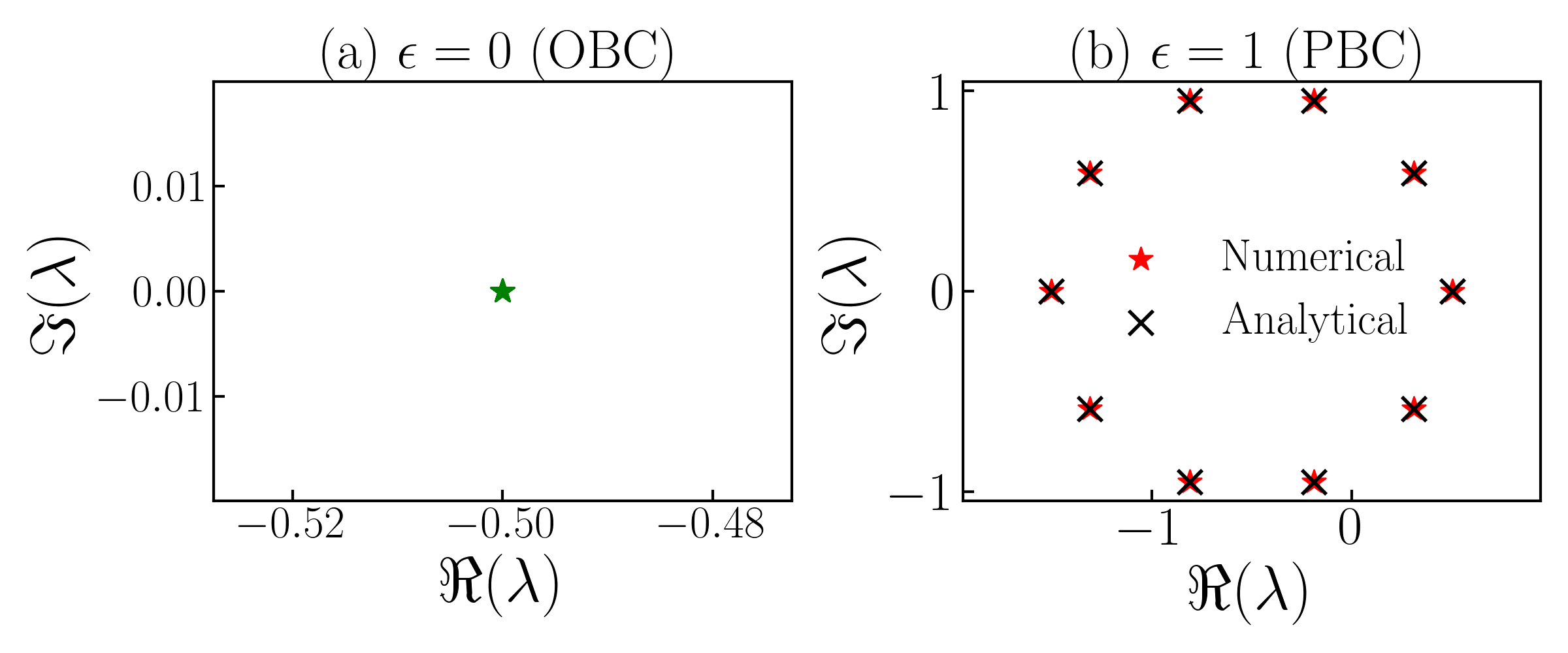}
    \caption{Model~III: We demonstrate in panel (a) the occurrence of an $N$-fold degeneracy of the eigenvalues of the drift matrix in Eq.~\eqref{eq:Model III-drift matrix} under OBC at the non-reciprocity coupling $g=0$. In contrast, PBC lifts this degeneracy, resulting in $N$ distinct eigenvalues, as shown in panel~(b). The parameters used are $k=-0.5$ and $N=10$.}
    \label{fig:EP-breaking-PBC}
\end{figure}

As mentioned before, the emergence of EP crucially depends on the specific choice of the open boundary condition (OBC) of the lattice. As for the periodic boundary condition (PBC), the drift matrix in Eq.~\eqref{eq:Model III-drift matrix} would be modified with additional non-zero elements $M_{1,N}=1$ and $M_{N,1}=g$. As a result, the EP at $g^*=0$ for the open boundary case gets completely destroyed. This stems from the fact that the boundary perturbation  to $J_N(0)$ of the form $V_{ij}^{\text{pert}} =\begin{cases}
    \epsilon,~(i,j) = (N,1) \\
    0,~ \text{otherwise}
\end{cases}  $ lifts the $N$ fold degeneracy for any $\epsilon >0$. Consequently the PBC drift matrix, $M(g=g^*,k) = -k\mathbb{I}_N + J_N(0) + V^{\text{pert}}(\epsilon=1)$ will split the otherwise $N$-fold degenerate eigenvalue $-k$ to $N$ distinct eigenvalues 
\begin{align}
    \{-k + \epsilon^{\frac{1}{N}} e^{2\pi im/N},~\forall m=1, \ldots,N\},
\end{align}
see Fig.~\ref{fig:EP-breaking-PBC}. 

\section{Anomalous relaxation: Correlation structure} \label{Sec:Anomalous relaxation}

In this section, we discuss the relaxation dynamics of the models discussed above for several quantities of interest, such as the autocorrelation function and the covariance matrix that includes cross-correlations. We also make some general remarks on multi-time multi-point correlators. To proceed, let us first define the autocorrelation function, 
\begin{align} \label{eq:defn-autocorrelation}
    \langle \vec{x}(0)^T \vec{x}(t)    \rangle \equiv  \vec{x}(0)^T e^{Mt}\vec{x}(0)    ,
\end{align}
where $\vec{x}(0)$ is the initial condition, and the equality above follows from $\langle \vec{x}(t) \rangle = e^{Mt} \vec{x}(0)$. In the steady state, the auto-correlation must be zero, i.e., $ \langle \vec{x}(0)^T \vec{x}(t)    \rangle_{\text{ss}} =0 $ since $ \langle \vec{x}(t) \rangle_{\text{ss}} =0$ from Eq.~\eqref{eq:coordiante soultion}. To capture the effect of non-reciprocity in a pronounced manner, we consider the relative autocorrelation function 
\begin{align} \label{eq:relative-autocor-defn}
    \mathcal{R}_{\text{Auto}}(g,t) = \frac{\langle \vec{x}(0)^T \vec{x}(t)    \rangle}  {\langle \vec{x}(0)^T \vec{x}(t)    \rangle_{\text{free}}},
\end{align}
where $\langle \vec{x}(0)^T \vec{x}(t)    \rangle_{\text{free}}$ is the corresponding autocorrelation function of the non-interacting model, namely, with the drift matrix elements $M_{ii} \neq 0$ and $M_{ij} =0, ~\forall~ i \neq j $.  

Next, defining the instantaneous fluctuation $\delta\vec{ x}(t) \equiv \vec{x}(t) - \langle \vec{x}(t) \rangle$, the covariance matrix is given by 
\begin{align}
    C (g,t) = \biggr\langle \delta\vec{ x}(t) \delta\vec{ x}^T(t) \biggr\rangle.
\end{align} 
Using the solution in Eq.~\eqref{eq:coordiante soultion} and the noise correlation, the covariance reduces to 
\begin{align} \label{eq:covariance-matrix} 
    C (g,t) = 2 D \int_0^t ds~ e^{M(g)s} e^{M^T(g)s},
\end{align}
and it further follows that the two-time correlator, $G(t,t')\equiv \langle \delta\vec{ x}(t) \delta\vec{ x}^T(t') \rangle $, can be expressed as 
\begin{align} \label{eq:Two-time-correlator}
    G(t,t')=e^{M(t-t')} C(g,t).
\end{align}
Moreover, any higher-order multi-time correlator can be written in terms of the two-point correlators following Wick's theorem~\cite{kardar2007}, e.g., the four-point multi-time correlator can be written as 
\begin{align}
    & \langle \delta x_i(t_1) \delta x_j(t_2) \delta x_k (t_3) \delta x_l(t_4) \rangle \nonumber \\
    & = C_{ij}(t_1,t_3) C_{kl}(t_2,t_4) 
    + C_{ik}(t_1,t_3) C_{jl}(t_3,t_4) \nonumber \\ & + C_{il}(t_1,t_4) C_{jk}(t_2,t_3).
\end{align}
Therefore, as expected for a Gaussian process, the equal-time covariance matrix suffices to characterize the entirety of all possible temporal and spatial fluctuations.

\subsection{Dynamics of Model I}

As discussed above, the dynamics of Model~I reduces to that of a single dimer with the corresponding drift matrix $ \widetilde{M} \equiv M^{(i,i+1)} $. We initiate the dynamics in the reduced space corresponding to a single dimer, with an initial condition  
$\vec{\tilde x}(0) = \begin{pmatrix}
    \cos{\theta} \\
    \sin{\theta}
\end{pmatrix}$. Without loss of generality, we set the initial value of the autocorrelation to unity, ensured by the above choice of the initial condition. From the structure of $\widetilde{M}$ in Eq.~\eqref{Eq:Block-decomposition-model I}, the autocorrelation function can be written as 
\begin{align}
    \langle \vec{\tilde x}(0)^T \vec{\tilde x}(t)    \rangle =  \vec{\tilde x}(0)^T  e^{\widetilde Mt}  \vec{\tilde x}(0)   =  \vec{\tilde x}(0)^T  \widetilde U e^{\Lambda t} \widetilde U^{-1} \vec{\tilde x}(0),
\end{align}
where $ \widetilde U = \begin{pmatrix}
        \sqrt{g} & -\sqrt{g} \\
        1 & 1
    \end{pmatrix}$
denotes the similarity transformation matrix with the inverse $\widetilde U^{-1} = (1/(2\sqrt{g})) \begin{pmatrix}
        1 & \sqrt{g} \\
        -1 & \sqrt{g}
    \end{pmatrix}$
such that $\widetilde M $ becomes diagonal, as $ \widetilde M =\widetilde U \Lambda \widetilde U^{-1} $ with $\Lambda = \begin{pmatrix}
        -k + \sqrt{g} & 0 \\
        0 & -k - \sqrt{g}
    \end{pmatrix}$. 
Following this, we obtain the closed-form expression of the autocorrelation function as
\begin{align} \label{eq:Auto-correlation-general-Model-I}
    \langle \vec{\tilde x}(0)^T \vec{\tilde x}(t)    \rangle = e^{-kt} \biggr[ \cosh{(\sqrt{g}t)} + \frac{\sin 2\theta}{\sqrt{g}} \sinh{(\sqrt{g}t)} \biggr].
\end{align}
This implies that the autocorrelation function decays with an exponential profile for all $g$ values; in particular, for $g\le 0$, the autocorrelation decays as $e^{-kt}$, and for $0<g < k^2$, the decay becomes faster according to the modified exponential $e^{-(k+g)t}$. To bring out features arising solely due to non-reciprocity, we probe the relative autocorrelation as defined in Eq.~\eqref{eq:relative-autocor-defn}, which yields the expression 
\begin{align}
    \mathcal{R}_{\text{Auto}}(g,t) = \cosh{(\sqrt{g}t)} + \frac{\sin 2\theta}{\sqrt{g}} \sinh{(\sqrt{g}t)}.
\end{align}
It then follows that for $g<0$, the quantity $\mathcal{R}_{\text{Auto}}(g,t)$ exhibits an oscillatory temporal profile, whereas for $0<g\leq k^2$, it displays exponential behavior.

\begin{figure}
    \centering
    \includegraphics[width=1\linewidth]{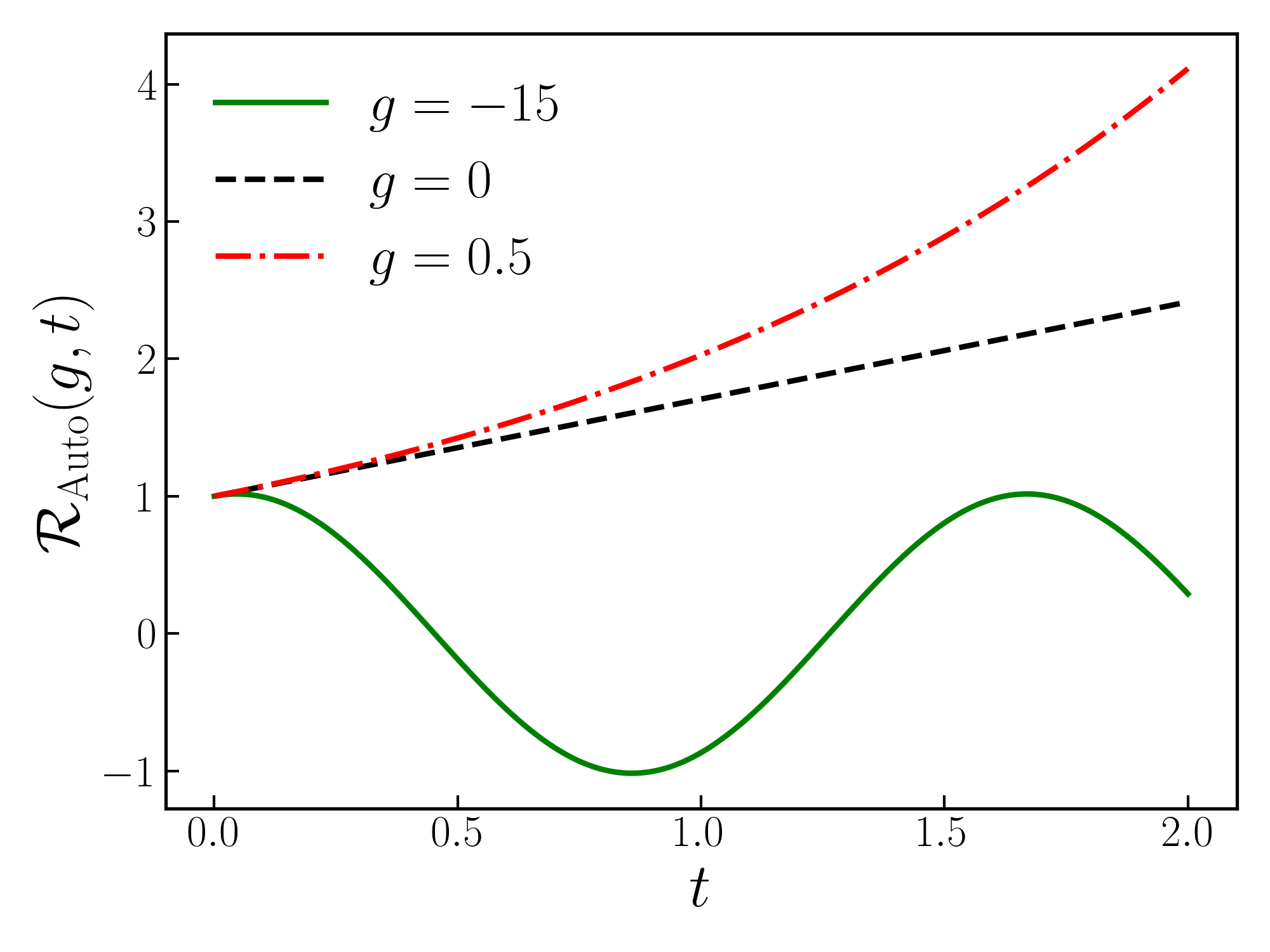}
    \caption{Model I: The behaviour of $\mathcal{R}_{\text{Auto}}(g,t)$ is shown for different representative values of $g$ in accordance with Eq.~\eqref{eq:relative-auto-model-I-behaviour}. The parameter values are $k=2,~\theta= \pi/8 $. The data used are based on exact results derived in Sec.~\ref{Sec:Anomalous relaxation}. }
    \label{fig:Relative-auto-Model-I}
\end{figure}

However, at the EP when $g=g^*=0$, the matrix $\widetilde U^{-1}$ is not well-defined, and the underlying Jordan-block structure renders quite remarkably the otherwise exponential decay to be dressed and hence, anomalous: one has an exponential decay modulated by a polynomial prefactor, yielding decay of the form $\sim e^{-kt} t$. In particular, using $\widetilde{M}(g=g^*) = \left[-k \mathbb{I}_2 + J_2(0)\right]$ gives us the exact expression of the autocorrelation function as
\begin{align} \label{eq:Autocorrelation-EP-Model I}
   &  \langle \vec{\tilde x}(0)^T \vec{\tilde x}(t)    \rangle = e^{-kt} (1 + t  \vec{\tilde x}_0^T J_2(0) \vec{\tilde x}_0) = e^{-kt} [1 + t \sin{2\theta}],\\
   & \mathcal{R}_{\text{Auto}}(g=g^*,t) = [1 + t \sin{2\theta}].
\end{align}
Indeed, taking the limit $g\to 0$ in the autocorrelation expression in Eq.~\eqref{eq:Auto-correlation-general-Model-I} reproduces correctly Eq.~\eqref{eq:Autocorrelation-EP-Model I}. Note that specific initial conditions with $\theta = 0, \pi/2, 3 \pi/2 $ evade anomalous relaxation, since for them the coefficient of polynomial dressing in Eq.~\eqref{eq:Autocorrelation-EP-Model I} becomes zero. These values of $\theta$ correspond to initial conditions that are either in the eigendirection of $J_2(0)$ or in its perpendicular direction. In summary, the quantity $\mathcal{R}_{\text{Auto}}$ will exhibit the following behavior:
\begin{align}\label{eq:relative-auto-model-I-behaviour}
    \mathcal{R}_{\text{Auto}}(g,t) \sim\begin{cases}
        &  \text{oscillatory};~g<0, \\
        &  \text{constant or linear growth};~g=g^*=0,\\
        &  \text{exponential growth};~0 < g <k^2, 
    \end{cases}
\end{align}
as captured in Fig.~\ref{fig:Relative-auto-Model-I}. 

Now, we discuss the corresponding situation for the covariance matrix $C(g,t)$. Using  
\begin{align}
    e^{\widetilde M s} = e^{-ks} \left[\cosh{(\sqrt{g}s)} \mathbb{I}_2 + \frac{\cosh{(\sqrt{g}s)}}{\sqrt{g} }Q \right]
\end{align} with $Q =\begin{pmatrix}
    0 & g \\
    1 & 0
\end{pmatrix}$ in Eq.~\eqref{eq:covariance-matrix}, we obtain 
\begin{widetext}
\begin{align}
    C(g,t) = 2D\int_{0}^t d s ~ e^{-2ks} \begin{pmatrix}
        \cosh^2{(s\sqrt{g})} + \frac{ \sinh^2{(s\sqrt{g})}}{g}& \frac{(1+g)}{\sqrt{g}} \cosh{(s\sqrt{g})}\sinh{(s\sqrt{g})}\\
        \frac{(1+g)}{\sqrt{g}} \cosh{(s\sqrt{g})}\sinh{(s\sqrt{g})} & \cosh^2{(s\sqrt{g})} + g \sinh^2{(s\sqrt{g})}
    \end{pmatrix}.
\end{align}
\end{widetext}
Clearly, all the elements have an exponential decay profile with the following steady-state behaviour:
\begin{align}
    C_{\text{ss}}(g) =\frac{D}{2k(k^2-g)} \begin{pmatrix}
        2k^2 -g & \frac{g+1}{2} \\
        \frac{g+1}{2} & 2k^2 - 1
    \end{pmatrix}.
\end{align}
We observe, somewhat counterintuitively, that at $g=-1$, the spatial cross-correlations between particles vanish at all times. The underlying mechanism and a generalized manifestation of this observation will be discussed later. Now, at the EP $g=g^*=0$, much like the autocorrelation function, the underlying Jordan canonical form dresses the exponential decay with a polynomial prefactor, yielding
\begin{widetext}
\begin{align}
    C(g=g^*,t)
    & = D \begin{pmatrix} \dfrac{1-e^{-2kt}}{k} & \dfrac{1-(1+2kt)e^{-2kt}}{2k^2} \\[2mm] \dfrac{1-(1+2kt)e^{-2kt}}{2k^2} & \dfrac{1-e^{-2kt}}{k} + \dfrac{1-(1+2kt+2k^2t^2)e^{-2kt}}{2k^3} \end{pmatrix}.
\end{align}
\end{widetext}
We remark that, in contrast to the autocorrelation in Eq.~\eqref{eq:Autocorrelation-EP-Model I}, the elements of the covariance matrix inevitably exhibit EP-induced anomalous relaxation, irrespective of the choice of initial condition.

\subsection{Dynamics of Model II}
Now, let us turn to Model~II, and investigate the effect of disorder. Utilizing the block diagonal structure as for Model~I, we analyze the dynamics in the reduced space of a single disordered dimer $(i,i+1)$  initialized as $\vec{\tilde x}(0) = \begin{pmatrix}
    \cos{\theta} \\
    \sin{\theta}
\end{pmatrix}$ with the drift matrix given by Eq.~\eqref{eq:Drfit-matrix-model-II}, 
\begin{align}
    \widetilde M \equiv M^{(i,i+1)} =  \begin{pmatrix}
        -k_i & g \\
        1 & -k_{i+1}
    \end{pmatrix}= \frac{(k_i + k_{i+1})}{2} \mathbb{I}_2 + R,
\end{align}
where $R = \begin{pmatrix}
    -\Delta k & g \\
    1 & \Delta k
\end{pmatrix}$ with $\Delta k= (k_i-k_{i+1})/2$. The matrix $R$ satisfies the relation $R^2 = \tilde \lambda^2 \mathbb{I}_2$ with $\tilde \lambda= \sqrt{(\Delta k)^2 + g }$, leading to 
\begin{align}
    e^{\widetilde M t } = e^{-\frac{k_i +k_{i+1}}{2}} \left[  \cosh{\tilde\lambda t} ~\mathbb{I}_2 + \frac{\sinh{\tilde \lambda t}}{\tilde \lambda} R \right].
\end{align}
This in turn gives us the following expression for the auto-correlation and the quantity $\mathcal{R}_{\text{Auto}}$ as 
\begin{widetext}
\begin{align}
    & \langle \vec{x}_0^T \vec{x}(t) \rangle  = e^{-\frac{k_i +  k_{i+1}}{2}t}
\left[
\cosh(\tilde\lambda t)
+
\frac{\sinh(\tilde \lambda t)}{\tilde \lambda}
\left(
-\Delta k\cos(2\theta)
+
\frac{g+1}{2}\sin(2\theta)
\right)
\right], \\
 & \mathcal{R}_{\text{Auto}} (g,t) = \left[
\cosh(\tilde \lambda t)
+
\frac{\sinh(\tilde \lambda t)}{\tilde \lambda}
\left(
-\Delta k\cos(2\theta)
+
\frac{g+1}{2}\sin(2\theta)
\right)
\right].
\end{align}
\end{widetext}
Now, as pointed out earlier, for a given realization of an individual dimer, the EP occurs exactly at $g^*=-(\Delta k)^2$, at which we have the following expressions: 
\begin{align}
    & \langle \vec{x}_0^T \vec{x}(t) \rangle  = e^{-\frac{k_i +  k_{i+1}}{2}t}
\left[
1
+
t
\left(
-\Delta k\cos(2\theta)
+
\frac{g+1}{2}\sin(2\theta)
\right)
\right], \\
& \mathcal{R}_{\text{Auto}} (g=g^*,t) = \left[
1
+
t
\left(
-\Delta k\cos(2\theta)
+
\frac{g+1}{2}\sin(2\theta)
\right)
\right].
\end{align}
Evidently, for a fixed realization $\{k_i\}$, we observe the same behaviour of   $\mathcal{R}_{\text{Auto}}$ as in Model~I, see Eq.~\eqref{eq:relative-auto-model-I-behaviour}, with one notable difference: at the EP, the polynomial dressing $\sim te^{-kt}$ is only absent for a single initial condition, namely, $\theta=0$.

Next, we analyze the elements of the covariance matrix in Eq.~\eqref{eq:covariance-matrix}. A straight-forward computation yields the following expressions of the elements:
\begin{align}
    & C_{11}(g,t) = 2D \int_0^t  \mathrm{d}t'~ e^{- (k_i + k_{i+1})t'} \biggr( c^2(t') \nonumber \\ 
     &~~~~~~~~~~~~~~~~~~~~~~ + \mathrm{s}^2(t') [ (\Delta k)^2 + g^2] - 2(\Delta k) c(t') \mathrm{s}(t')\biggr),  \\
     & C_{22}(g,t) = 2D \int_0^t  \mathrm{d}t'~ e^{- (k_i + k_{i+1})t'}\biggr( c^2(t') \nonumber \\ 
     &~~~~~~~~~~~~~~~~~~~~~~~~~~~~~~ \mathrm{s}^2(t') [ (\Delta k)^2 + 1] + 2(\Delta k) c(t') \mathrm{s}(t')\biggr), \\
     & C_{12}(g,t)  = 2D \int_0^t  \mathrm{d}t'~ e^{- (k_i + k_{i+1})t'}\biggr(  \mathrm{s}^2(t') (\Delta k) [ 1 -g ] \nonumber \\ 
     &~~~~~~~~~~~~~~~~~~~~~~~~~~~~~~~~~~~~~~~~~~~~~~~~+ (1+g) c(t') \mathrm{s}(t')\biggr) \nonumber \\
     &  ~~~~~~~~~~~~~~~= C_{21}(g,t),
\end{align}
where $c(t')\equiv \cosh{(\tilde\lambda t')} $ and $\mathrm{s}(t') \equiv \sinh{(\tilde \lambda t')}/\tilde\lambda$. The steady-state covariance matrix takes the form
\begin{align}
    C_{\text{ss}}(g) & = \frac{D}{K(k_ik_{i+1}-g)} \nonumber \\
& ~~\times \begin{pmatrix}
k_{i+1}K+g(g-1)
&
k_{i+1}+gk_i
\\[4pt]
k_{i+1}+gk_i
&
k_iK+1+g
\end{pmatrix},
\end{align}
where $K\equiv k_i + k_{i+1}$.

At EP with $g= g^* = -(\Delta k)^2$, the full covariance matrix is 
\begin{widetext}
    \begin{align}
C(g=g^*, t)
=
\begin{pmatrix}
\displaystyle
\frac{2D}{K}F_0
-\frac{4D(\Delta k)}{K^2}F_1
+\frac{4D\left((\Delta k)^2+(\Delta k)^4\right)}{K^3}F_2
&
\displaystyle
\frac{2D\left(1-(\Delta k)^2\right)}{K^2}F_1
-\frac{4D\left((\Delta k)+(\Delta k)^3\right)}{K^3}F_2
\\[12pt]
\displaystyle
\frac{2D\left(1-(\Delta k)^2\right)}{K^2}F_1
-\frac{4D\left((\Delta k)+(\Delta k)^3\right)}{K^3}F_2
&
\displaystyle
\frac{2D}{K}F_0
+\frac{4D(\Delta k)}{K^2}F_1
+\frac{4D\left(1+(\Delta k)^2\right)}{K^3}F_2
\end{pmatrix},
\end{align}
\end{widetext}
where 
\begin{align}
    & F_0 (t)= 1 - e^{-Kt},\\
    & F_1 (t)= 1- ( 1 + K)t  e^{-Kt}, \\
    &  F_2 (t)= 1- \left[1 + Kt + \frac{K^2t^2}{2!} \right] e^{-Kt}. 
\end{align} 

\subsection{Dynamics of Model III}
Let us now write down the propagator of the drift matrix in Eq.~\eqref{eq:Model III-drift matrix} for Model~III. In particular, we are interested in the matrix elements $(e^{M(g)t})_{ij}$. In order to achieve this, we first note the following: 
\begin{align}
    M(g) = -k \mathbb{I}_{N} + T(g),
\end{align}
where the asymmetric tri-diagonal matrix $T(g)$ can be mapped to a symmetric tri-diagonal matrix $A$ as 
\begin{align}
    T(g) = \mathcal{D} ( \sqrt{g} A )\mathcal{D}^{-1}, 
\end{align}
where $(\mathcal{D})_{ij} = g^{(i-1)/2} \delta_{ij}$, and $A_{ij} = \delta_{i,j-i-1}$ and $A_{ij}= A_{ji}$ with $1 \le i,j \le N$, and $\delta_{ij}$ is as usual the Kronecker delta. Indeed, the transformation matrix $\mathcal{D}$ corresponds to a non-unitary similarity transformation, i.e., 
$
\mathcal{D}^\dagger \mathcal{D} \neq \mathbb{I}_N, ~ \mathcal{D}^{-1} \mathcal{D}= \mathcal{D} \mathcal{D}^{-1} =\mathbb{I}_N.  
$
This transformation to the symmetric matrix $\sqrt{g}A$ simplifies the computation of the propagator, since $A$ can be interpreted as the Hamiltonian of the tight-binding model  with open boundary conditions,  
\begin{align}\label{eq:tight-binding Hamiltonian}
    A = \sum_{i=1}^{N-1} \left(| i \rangle \langle i+1| + | i  +1 \rangle \langle i | \right), 
\end{align}
for which the eigenvalues and eigenfunctions are known exactly. Using this mapping, we obtain (for details, see Appendix~\ref{APP:Derivation-exponential-Hatano-Nelson}) the following expression for the propagator:
\begin{align} \label{eq:Exponential-matrix-elements}
    \left(e^{M(g)t} \right)_{ij} = & e^{-kt} g^{(i-j)/2} \frac{2}{N+1} \nonumber \\
     & \times \sum_{m=1}^N e^{2\sqrt{g} t \cos(\theta_m)} \sin{(i \theta_m )} \sin{(j \theta_m )} ,
\end{align}
where $\theta_m = m \pi /(N+1)$. 

Equation~\eqref{eq:defn-autocorrelation} gives 
\begin{align}
    \langle \vec x^T(0) e^{Mt} \vec x(0)\rangle = \sum_{ij=1}^N x_i(0)  \left(e^{M(g)t} \right)_{ij} x_j(0),
\end{align}
which upon using Eq.~\eqref{eq:Exponential-matrix-elements} yields
\begin{align}
    & \langle \vec x^T(0) e^{Mt} \vec x(0)\rangle \nonumber \\
    & =\frac{2 e^{-kt} }{N+1} \sum_{m=1}^N e^{2\sqrt{g} t \cos(\theta_m)} \biggr[\sum_{i=1}^N x_i(0) g^i \sin{(i \theta_m )} \biggr] \nonumber \\
    & ~~~~~ \times \biggr[\sum_{i=1}^N x_j(0) g^{-j} \sin{(j \theta_m )} \biggr].
\end{align}
Defining the projectors $X_m^{\pm }$ that obtain the overlap of the initial vector with the eigenvectors of the tight-binding Hamiltonian, as 
\begin{align}
    X_m^{\pm } \equiv \sum_{i=1}^N  x_i(0)~ g^{\pm i}~ \sin{(i \theta_m )},
\end{align}
we write down the auto-correlation function in the following compact form: 
\begin{align}
     \langle \vec x^T(0) e^{Mt} \vec x(0)\rangle =\frac{2 e^{-kt} }{N+1} \sum_{m=1}^N X_m^+ X_m^- ~  e^{2\sqrt{g} t \cos(\theta_m)},
\end{align}
and the quantity $\mathcal{R}_{\text{Auto}}$ takes the form 
\begin{align}
    \mathcal{R}_{\text{Auto}}(g,t) = \frac{2}{N+1} \sum_{m=1}^N X_m^+ X_m^- ~  e^{2\sqrt{g} t \cos(\theta_m)}.   
\end{align}

However, at the exceptional point when $g=g^*=0$, the expression must be modified to account for the underlying Jordan canonical form~\eqref{eq:Jordan-decomposition-Model-III} of the drift matrix. We have
\begin{align}
    e^{M(g^*=0) t} = e^{-kt} e^{ J_N(0) t } = e^{-kt} \sum_{p=0}^{N-1} \frac{(J_N(0))^p}{p!} t^p,
\end{align}
where we have used the nilpotency relation $(J_N(0))^N =0$. Now, noting that $[J_N(0)]_{ij} = \delta_{i,j-1}$, we obtain 
\begin{align}
    [(J_N(0))^p]_{ij} & = \sum_{k_1, k_2, \ldots, k_{p-1}=1}^N (J_N(0))_{ik_1} (J_N(0))_{k_1k_2}\ldots \nonumber \\
    & ~~~~~~~~~~~~~~~~~~~ \times (J_N(0))_{k_{p-1}j } \nonumber \\
    & = \sum_{k_1, k_2, \ldots, k_{p-1}=1}^N \delta_{i, k_1-1} \delta_{k_1, k_2-1}\ldots \delta_{k_{p-1}, j-1} \nonumber \\ 
    & = \sum_{k_2, \ldots, k_{p-1}=1}^N  \delta_{i+1, k_2-1} \delta_{k_2, k_3-1}\ldots \delta_{k_{p-1}, j-1} \nonumber \\
    & \vdots \nonumber \\
    & = \sum_{k_{p-1}=1}^N \delta_{ i+ p-1 , k_{p-1}}\delta_{k_{p-1}, j-1} \nonumber \\
    & = \delta_{i+p,j}.
\end{align}
This gives us 
\begin{align}
    \left[e^{M(g^*=0) t}\right]_{ij} & = e^{-kt} \sum_{p=0}^{N-1} \frac{t^p }{p!} [(J_N(0))^p]_{ij} \nonumber \\
    & = e^{-kt} \sum_{p=0}^{N-1} \frac{t^p }{p!} \delta_{i+p,j} \nonumber \\
    & = \begin{cases}
        e^{-kt} \frac{t^{j-i}}{(j-i)!},~ j\ge i \\
        0 ,~~~~~~~~~~ j<  i.
    \end{cases} \label{eq:Exponential-elements-EP-model-III}
\end{align}
Combining these results, we obtain the following expression of the autocorrelation function at EP:
\begin{align}
    \langle \vec x^T(0) e^{M(g=g^*)t} \vec x(0) \rangle & = e^{-kt} \sum_{p=0}^{N-1} \frac{t^p }{p!} \sum_{i,j=1}^N x_i(0) x_j(0)   \delta_{i+p,j} \nonumber \\
    & = e^{-kt} \sum_{p=0}^{N-1} \frac{t^p }{p!} \sum_{i=1}^{N-p}  x_i(0) x_{i+p}(0),
\end{align}
which evidently contains the following polynomial dependence in the autocorrelation and in $\mathcal{R}_{\text{Auto}}$ at $g=g^*=0$,
\begin{widetext}
\begin{align}
    & \langle \vec x^T(0) e^{M(g=g^*)t} \vec x(0) \rangle = e^{-kt} \biggr[  \sum_{i=1}^N x_i^2(0) + t \sum_{i=1}^{N-1} x_i(0) x_{i+1}(0) + \frac{t^2}{2!} \sum_{i=1}^{N-2} x_i(0) x_{i+2}(0) + \ldots + \frac{t^{N-1}}{(N-1)!} x_1(0) x_N(0) \biggr], \\
    & \mathcal{R}_{\text{Auto}} (g=g^*,t)= \biggr[  \sum_{i=1}^N x_i^2(0) + t \sum_{i=1}^{N-1} x_i(0) x_{i+1}(0) + \frac{t^2}{2!} \sum_{i=1}^{N-2} x_i(0) x_{i+2}(0) + \ldots + \frac{t^{N-1}}{(N-1)!} x_1(0) x_N(0) \biggr]. \label{eq:Relative-auto-EP-model-III}
\end{align}
\end{widetext}
Evidently, the polynomial degree observed in $\mathcal{R}_{\text{Auto}}$ depends explicitly on the initial condition. Indeed, the coefficient of the $t^m$ term in Eq.~\eqref{eq:Relative-auto-EP-model-III} is given by $\sum_{i=1}^{N-m} x_i(0)x_{i+m}(0)$, and may vanish for special choices of $\vec{x}(0)$. In particular, the maximally allowed polynomial degree of $(N-1)$ is present only when $x_1(0)x_N(0)\neq 0$. Moreover, starting from a localized initial condition $x_{i}(0)= \delta_{i,i_0}$ evades the polynomial dependence entirely, and $\mathcal{R}_{\text{Auto}} =1$ at all times. Thus, the EP determines the maximal polynomial degree permitted by the Jordan structure, while the actual polynomial dependence is controlled by the choice of the initial condition. In summary, a similar characterization as in Eq.~\eqref{eq:relative-auto-model-I-behaviour} for Model~I is also deducible: 
\begin{align}
        \mathcal{R}_{\text{Auto}} (g,t) \sim\begin{cases}
        &  \text{oscillatory};~g<g^*, \\
        &  \text{ $t^{\alpha}$ with $0\le \alpha\le N-1$ };~g=g^*,\\
        &  \text{exponential growth};~g<k/2. 
    \end{cases}
\end{align}

Let us now delve into the analysis of the covariance matrix, which from Eq.~\eqref{eq:covariance-matrix} in conjunction with Eq.~\eqref{eq:Exponential-matrix-elements} follows 
\begin{widetext}
\begin{align}
    [C(g,t)]_{ij} & = 2D \int_0^t \mathrm{d}t'~ \sum_{l=1}^N (e^{M(g) t' })_{il} (e^{M^T(g) t' })_{lj} \nonumber \\
    & = \frac{8D g^{(i+j)/2}}{(N+1)^2}  \sum_{m,n=1}^N \biggr(\int_0^t \mathrm{d}t'~e^{-2kt' + 2\sqrt{g} t' [\cos(\theta_m) + \cos(\theta_n)] }\biggr) ( \sin{(i\theta_m)} \sin{(j\theta_n)} ) \left( \sum_{l=1}^N g^{-l} \sin{(l\theta_m)} \sin{(l\theta_n)} \right) \nonumber \\ 
    & = \frac{8D g^{(i+j)/2}}{(N+1)^2}  \sum_{m,n=1}^N F_{mn}(g) \sin{(i\theta_m)} \sin{(j\theta_n)} \left(\frac{1 - e^{\Lambda_{mn}t}}{\Lambda_{mn}}\right)\label{eq:Covariance-model-III},
\end{align} 
\end{widetext}
where we have used $\Lambda_{mn}\equiv 2k - 2\sqrt{g}  [\cos(\theta_m) + \cos(\theta_n)] $, and $F_{mn}(g) \equiv \sum_{l=1}^N g^{-l} \sin{(l\theta_m)} \sin{(l\theta_n)}$. The steady-state covariance is therefore simply 
\begin{align} \label{eq:steady-state-covariance-model-III}
    [C_{\text{ss}}(g)]_{ij} & = \frac{8D g^{(i+j)/2}}{(N+1)^2}  \sum_{m,n=1}^N \frac{F_{mn}(g)}{\Lambda_{mn}} \sin{(i\theta_m)} \sin{(j\theta_n)}. 
\end{align}

\begin{figure}
    \centering
    \includegraphics[width=1\linewidth]{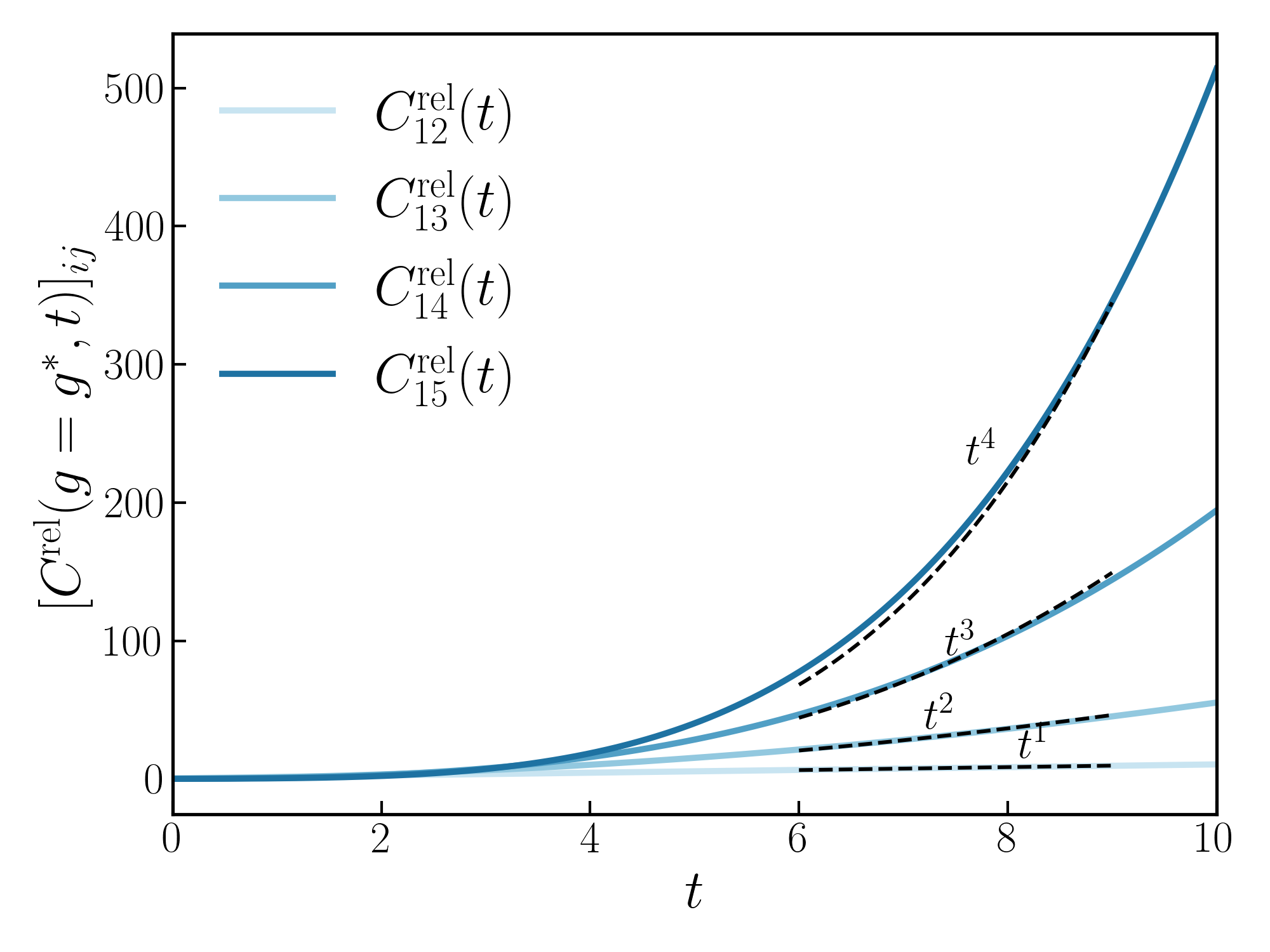}
    \caption{Model II: The temporal behavior of the relative covariance matrix elements defined in Eq.~\eqref{eq:relative covariance} is shown. The dominant power-law dependence on time is indicated by the black dashed lines. The parameters are $k=D=1$. 
    The data used here are based on exact results derived in Sec.~\ref{Sec:Anomalous relaxation}.}
    \label{fig:Polynomial-dependence_Cov_Model_III}
\end{figure}

At EP, when $g=g^*=0$, using Eq.~\eqref{eq:Exponential-elements-EP-model-III}, we obtain the modified covariance as 
\begin{align}\label{Eq:Cross-correlation-temporal-profile-Model-III}
    C_{ij}(g=g^*,t)& = 2D \sum_{l=1}^{\min(i,j)} \frac{1}{(i-l)!(j-l)!} \nonumber \\
    &~~~~~~~\times\int_{0}^t \mathrm{d}t' ~ e^{-2kt'} (t')^{i+j-2l}.
\end{align}
The above expression is highly non-intuitive in the sense that the cross-correlation relaxes with a polynomial dependence whose degree explicitly depends on the spatial distance between the particles for which the cross-correlation is being measured. In other words, the temporal profile of the relaxation of the covariance encodes information about the underlying spatial structure of the lattice: the dynamics itself reveals the spatial separation and location of the particles whose correlations are being measured. We also note that the steady-state covariance matrix allows a particularly neat expression:
\begin{align}
    [C_{\text{ss}}(g=g^*)]_{ij} = 2D
    \sum_{l=1}^{\min(i,j)}
    \frac{(i+j-2l)!}
    {(i-l)!(j-l)!(2k)^{i+j-2l+1}}. 
\end{align}
An efficient way to observe the polynomial relaxation of covariance elements due to EP is to analyze the following relative quantity:
\begin{align}
    [C^{\text{rel}} (g=g^*,t)]_{ij} & \equiv e^{2kt} [C_{\text{ss}}(g=g^*) - C(g=g^*,t) ]_{ij} \nonumber \\
    & = 2D
    \sum_{l=1}^{\min(i,j)}
    \frac{e^{2kt}}{(i-l)!(j-l)!} \nonumber \\
    &~~~~~~~~\times \biggr(\int_t^\infty \mathrm{d}t'~
    e^{-2kt'}(t')^{i+j-2l} \biggr),
\end{align}
which, upon using the identity \begin{align}
    \int_t^\infty \mathrm{d}t'\, e^{-2kt'}(t')^n
    &=
    \frac{n!}{(2k)^{n+1}}e^{-2kt}
    \sum_{r=0}^{n}\frac{(2kt)^r}{r!},
\end{align}
yields
\begin{align} \label{eq:relative covariance}
    [C^{\text{rel}} (g=g^*,t)]_{ij}
    &=
    2D
    \sum_{l=1}^{\min(i,j)}
    \frac{(i+j-2l)!}
    {(i-l)!(j-l)!(2k)^{i+j-2l+1}} \nonumber \\
    &~~~~~~~~~~~~~~\times  \biggr(\sum_{r=0}^{i+j-2l}
    \frac{(2kt)^r}{r!}\biggr).
\end{align}
The last expression is particularly useful since it eliminates the exponential damping present in the original covariance matrix, and lets one clearly observe only the polynomial time dependence, as demonstrated in Fig~\ref{fig:Polynomial-dependence_Cov_Model_III}.


\section{Pseudo-equilibrium at points of complete asymmetry} \label{Sec:Pseudo-Equilibrium}

We now discuss a remarkable phenomenon that occurs in Model III (also in Model I) when the non-reciprocity parameter is set to $g=-1$. At this specific value, the drift matrix $M$  in Eq.~\eqref{eq:Model III-drift matrix} shows chirality since it anti-commutes with the matrix $\Gamma_{ij}= (-1)^{i} \delta_{ij} $, i.e., $\{M, \Gamma\}=0$, and it can be decomposed as $M(g=-1,k) = -k \mathbb{I}_N + A_N $, where $A_N = - A_N^T $ is an anti-symmetric tridiagonal matrix with entries in the upper diagonal being $+1$, and those in the lower diagonal being $-1$. We now show that a significant simplification of the covariance structure emerges at $g=-1$, which is not immediately apparent from Eq.~\eqref{eq:Covariance-model-III}. Computing the covariance directly from Eq.~\eqref{eq:covariance-matrix}, we obtain
\begin{align}
C(g=-1,t)
&= 2D\mathbb{I}_N\int_0^t ds\,e^{-2ks} \nonumber\\
&= \frac{D}{k}\left(1-e^{-2kt}\right)\mathbb{I}_N.
\end{align}
Thus, despite the presence of non-reciprocal interactions, the dynamics does not generate any equal-time cross-correlations. However, the two-time correlator in Eq.~\eqref{eq:Two-time-correlator} does encode nontrivial cross-correlations, as do higher-order correlators. Nevertheless, the complete absence of equal-time cross-correlations suggests ``dynamical factorization'' at $g=-1$, whereby the non-reciprocal interaction remains present in the dynamics but is invisible to equal-time observables. This gives the system an effectively ``non-interacting'' character at the level of equal-time statistics. In the following, we unveil the mathematical origin of this remarkable behavior.

For any antisymmetric matrix $A_N$, one can find a rotation matrix $Q$ such that in the rotated frame, the following canonical block decomposition holds~\cite{Bowen2008}:
\begin{align} \label{eq:Antisymmetric-canonicl form}
    Q^TA_NQ = \bigoplus_{j=1}^{N/2} \begin{pmatrix}
        0 & \omega_j \\
        - \omega_j & 0
    \end{pmatrix},
\end{align}
where $\omega_j$'s are singular values of $A_N$ and $Q^TQ= Q Q^T= \mathbb{I}_N$. Note that for our case, $\omega_j$'s  are obtained from  Appendix~\ref{APP:Hatano-Nelson-Spectrum} by by putting $g=-1$, yielding $\omega_j = |E_m| = 2 \cos\theta_m $.  
\subsection{Dynamical reduction}
Now, we investigate the structure of the OU dynamics in the rotated frame. Let us define $\vec y \equiv Q^T \vec x$, for which the dynamics $\dot{ \vec{y}} = Q^T \dot{\vec x}$ becomes on using Eq.~\eqref{eq:multi-variate OU} that 
\begin{align}
    \dot{ \vec{y}} = \left[\bigoplus_{j=1}^{N/2} \begin{pmatrix}
        -k & \omega_j \\
        -\omega_j & -k 
    \end{pmatrix}\right] \vec y   + \vec\xi(t),  
\end{align}
where $\vec\xi(t) = Q^T \vec \eta(t) $ with $\langle \vec \xi(t) \vec \xi^T (t') \rangle = 2D \mathbb{I}_N \delta(t - t') $. Evidently, this decomposition reduces the $N$-particle dynamics to $N/2$ numbers of disjoint $2$-particle dynamics, with the following equation of motion
\begin{align}
    \begin{pmatrix}
        \dot y_{2j-1}\\
        \dot y_{2j}
    \end{pmatrix} = \begin{pmatrix}
        -k & \omega_j \\
        -\omega_j & -k 
    \end{pmatrix} \begin{pmatrix}
         y_{2j-1}\\
         y_{2j}
    \end{pmatrix} + \begin{pmatrix}
        \xi_{2j-1}\\
        \xi_{2j}
    \end{pmatrix}.
\end{align}
Introducing the complex  variables $z_j= y_{2j-1} + i y_{2j}$, each of the two-dimensional OU processes can be written as  
\begin{align} \label{eq:complex-Gaussian}
    \dot z_{j} = (-k -i \omega_j) z_j + \zeta_j(t),
\end{align}
where $\zeta_j(t)= \xi_{2j-1} + i \xi_{2j}  $ is a complex Gaussian white noise with correlation $\langle \zeta_j(t) \zeta_{j'}^*(t') \rangle = 4D \delta_{jj'} \delta(t-t') $. Here, * denotes complex conjugation. The process~\eqref{eq:complex-Gaussian} has the steady-state distribution 
\begin{align}
    P_{\text{ss}}(z_j,z_j^*) = \frac{k}{2\pi D} \exp{\left[\frac{- k |z_j|^2}{2 D} \right]},
\end{align}
which in the $\{y_{2j-1},y_{2j}\}$ coordinate, following probability conservation and $dy_{2j-1}dy_{2j} = 2~ dz_j dz_j^*$, becomes 
\begin{align}
    P_{\text{ss}}(y_{2j-1}, y_{2j}) = \frac{k}{\pi D} \exp{\left[\frac{- k (y_{2j-1}^2 + y_{2j}^2)  }{2 D} \right]}. 
\end{align}
Therefore, the steady-state measure of the entire system in the rotated frame becomes 
\begin{align}
    P_{\text{ss}}(\{y_j\}) = \left(\frac{k}{\pi D}\right)^{N/2} ~\exp{\left[- \frac{ k   }{2 D} \vec y^T \vec y \right]}.
\end{align}

Since the original coordinate $\vec{x}$ is related to $\vec y$ via rotation by an orthogonal matrix, we have $\vec y^T \vec y = \vec x^T \vec x$. Hence, the steady-state measure of the entire system retains the same form:
\begin{align}
    P_{\text{ss}}(\{x_j\}) & = \left(\frac{k}{\pi D}\right)^{N/2} ~\exp{\left[- \frac{ k \sum_{j=1}^N x_{j}^2   }{2 D} \right]} \nonumber \\
    & = \prod_{j=1}^N \sqrt{\frac{k}{\pi D}} ~\exp{\left[- \frac{ k x_{j}^2   }{2 D} \right]}.
\end{align}
This factorization has a direct physical interpretation. Despite the presence of non-reciprocal interactions between neighboring particles, the steady-state distribution contains no static correlations between different particle coordinates. In particular,
\begin{align} \label{eq:anti-symmetric-covariance}
\langle x_i x_j\rangle_{\mathrm{ss}}
=\frac{D}{k}\delta_{ij},
\end{align}
so that the equal-time fluctuations of different particles are statistically independent. Each particle therefore has exactly the same marginal steady-state distribution as that of an isolated Ornstein--Uhlenbeck particle trapped in a harmonic potential $V(x)=kx^2/2$ and subjected to Gaussian white noise of strength $D$. However, this does not mean that the steady state of the system corresponds to equilibrium. The non-reciprocal coupling remains present in the equations of motion, giving rise to non-zero steady-state probability currents in the phase space, signaling non-equilibrium as we show below; consequently, we have a steady state that is only ``pseudo-equilibrium." It is well known~\cite{Risken1989} that the steady-state current for Eq.~\eqref{eq:multi-variate OU} is
\begin{align}
    \vec{J}_{\mathrm{ss}}(\vec{x})
    =
    [M+DC_{\mathrm{ss}}^{-1}]\vec{x}\,P_{\mathrm{ss}}(\vec{x}).
\end{align}
At $g=-1$, using $C_{\mathrm{ss}}=(D/k)\mathbb{I}_N$ from Eq.~\eqref{eq:anti-symmetric-covariance}, we get
\begin{align}
    \vec{J}_{\mathrm{ss}}(\vec{x})
    =
    A_N\vec{x}\,P_{\mathrm{ss}}(\vec{x}).
\end{align}
Thus, although the steady-state measure retains its equilibrium form, the current is nonzero. Moreover, since $\vec{x}^{T}A_N\vec{x}=0$,
\begin{align}
    \vec{J}_{\mathrm{ss}}(\vec{x})\cdot\vec{x}=0,
\end{align}
implying that the current is purely tangential to hyperspheres of fixed radius $\vec{x}^{T}\vec{x}$ in the state space.

Our model provides an explicit realization of the general framework discussed in Ref.~\cite{Kwon2005}, where a Gibbs--Boltzmann steady-state measure can coexist with a nonzero circulating current. In particular, the stochastic dynamics of the form
\begin{align}
    \frac{d\vec{x}}{dt}
    =
    -\vec{\nabla}U(\vec{x})
    +\vec{F}_{\mathrm{nc}}(\vec{x})
    +\vec{\eta}(t),
\end{align}
where the non-conservative force satisfies
\begin{align}
    \vec{\nabla}\cdot\vec{F}_{\mathrm{nc}}(\vec{x})=0,
    \qquad
    \vec{F}_{\mathrm{nc}}(\vec{x})\cdot\vec{\nabla}U(\vec{x})=0,
\end{align}
leads to the discussed situation.
For our model, $\vec{F}_{\mathrm{nc}}(\vec{x})=A_N\vec{x}$ and $U(\vec{x})=k\vec{x}^{T}\vec{x}/2$; additionally, the linear antisymmetric dynamics does not generate any equal-time cross-correlations at any time.

\section{Heat dissipation and fluctuation-dissipation theorem violation metric} \label{Sec:Heat-dissipation}

A natural way to characterize the departure of a steady state from equilibrium is to examine the relation between its spontaneous fluctuations and its response to weak external perturbations. Linear-response theory provides a framework for establishing this connection: the correlation function characterizes the fluctuations generated by the unperturbed dynamics, while the response function quantifies the change in the system induced by an infinitesimal perturbation. At equilibrium, these two quantities are not independent but are related by the fluctuation-dissipation theorem (FDT), which follows from detailed balance and the resulting canonical equilibrium distribution. An NESS, in contrast, can possess well-defined correlations and responses while violating this equilibrium relation. The FDT therefore provides a natural equilibrium benchmark, and its violation offers an operational measure of the departure from equilibrium. For multidimensional Ornstein-Uhlenbeck processes, the corresponding fluctuation-response relations can be formulated explicitly in terms of the drift, diffusion, correlation, and response matrices. 

To discuss the FDT in our setup, we introduce a weak time-dependent additive perturbation $\vec{h}(t)$ through
\begin{align}
    \dot{\vec{x}}
    =
    M\vec{x}
    +
    \vec{\eta}(t)
    +
    \vec{h}(t),
    \label{eq:perturbed OU}
\end{align}
and define the response matrix as
\begin{align}
    R_{ij}(t,t')
    \equiv
    \left.
    \frac{\delta\langle x_i(t)\rangle}
    {\delta h_j(t')}
    \right|_{\vec h=0},
    \label{eq:response-definition}
\end{align}
while the two-time correlation matrix is
\begin{align}
    G_{ij}(t,t')
    =
    \left\langle x_i(t)x_j(t')\right\rangle .
    \label{eq:correlation-definition}
\end{align}
At steady state, both the quantities depend only on the time difference, i.e., $R_{ij} (t,t') \equiv R_{ij}(\tau)$ and $G_{ij} (t,t') \equiv G_{ij}(\tau)$, where $\tau\equiv t-t'$. The corresponding frequency-space representations via Fourier transformation are $\widetilde R(\omega)$ and $\widetilde G(\omega)$, respectively. In the frequency space, the equilibrium FDT takes the form
\begin{align}
    \widetilde G(\omega)
    =
    \frac{2D}{\omega}
    \Im(\widetilde R(\omega)),
    \label{eq:FDT}
\end{align}
where $\Im(\widetilde R(\omega)) \equiv ( \widetilde R(\omega) - \widetilde R(\omega)^\dagger )/(2i)$ denotes the anti-Hermitian part of the response matrix~\cite{Kubo_1966,Marconi2008}. The derivation of the response and correlation functions for the multidimensional OU process, together with their frequency-domain representations, is provided in Appendix~\ref{APP:FDT derivation}. We find that 
\begin{align}
    \tilde{R}(\omega) & = [-i\omega - M]^{-1}, \label{eq:Response}  \\\tilde{G}(\omega) & = 2D [-i\omega - M]^{-1} [i\omega - M^T]^{-1} \nonumber \\
    & = 2D \tilde{R}(\omega) \tilde{R}^\dagger(\omega). \label{eq:Correlation}
\end{align} 

At NESS, Eq.~\eqref{eq:FDT} is generically violated. We therefore define the FDT-violation matrix as
\begin{align}
    \chi(\omega)
    \equiv
    \widetilde G(\omega)
    -
    \frac{2D}{\omega}
    \Im(\widetilde R(\omega)).
    \label{eq:FDT-violation}
\end{align}
By construction, $\chi(\omega)$ vanishes at equilibrium when the FDT is satisfied, while a finite $\chi(\omega)$ quantifies the departure from the FDT. 

A remarkable equality known as the Harada-Sasa relation~\cite{Harada2005} provides a direct link between the FDT violation matrix for the velocity $\vec v= \dot {\vec x}$ and the heat dissipated into the heat bath. This provides a natural framework for quantifying the heat dissipation of individual particles arising from non-reciprocal interactions across our hierarchy of models. In particular, the Harada--Sasa relation yields the heat dissipation of the $i$'th particle as
\begin{align}
    \langle Q_i \rangle = \int_{-\infty}^{\infty} \frac{\mathrm{d}\omega}{2\pi} \, \left[ \tilde G^v_{ii}(\omega) -2 D \Re[\tilde R^v_{ii }(\omega)] \right],  
\end{align} 
and the total heat dissipation to the bath would therefore be $\langle Q_{\text{tot}}\rangle =\sum_{i} \langle Q_i \rangle$. Now, let us note that $\vec v(\omega) = -i\omega \vec x(\omega) $, leading to $\tilde R^v(\omega) = - i\omega R^x(\omega)$, and $\tilde G^v(\omega) = \langle \vec v(\omega) \vec v^\dagger (\omega) \rangle = \omega^2 \tilde G(\omega) $. This, in turn, gives using Eq.~\eqref{eq:FDT-violation} that  
\begin{align}
    \langle Q_i \rangle & =  \int_{-\infty}^{\infty} \frac{\mathrm{d}\omega}{2\pi} \,\omega^2 \, \chi_{ii} (\omega), \label{eq:Individual-dissipated-heat} \\
    \langle Q_{\text{tot}}\rangle & = \int_{-\infty}^{\infty} \frac{\mathrm{d}\omega}{2\pi} \,\omega^2 \text{Tr}[\chi]. 
\end{align}

\subsection{For Model~I}
Now, for the drift matrix in Eq.~\eqref{Eq:Block-decomposition-model I} in the reduced space of a single dimer, we obtain the response matrix in frequency domain from Eq.~\eqref{eq:Response} as 
\begin{align}
    \tilde R(\omega)= \frac{1}{\Delta(\omega)} \begin{pmatrix}
        k-i\omega & 1 \\
        g & k-i\omega
    \end{pmatrix},
\end{align}
where $\Delta(\omega) = (k-i\omega)^2 -g$. Similarly, the correlation matrix from Eq.~\eqref{eq:Correlation} is obtained as 
\begin{align}
    \tilde G(\omega) & = \frac{2D}{|\Delta(\omega)|^2 } \nonumber \\ & ~~~~\times \begin{pmatrix}
         k^2 + \omega^2 + 1& (k-i\omega) + g (k+i\omega) \\
        (k+i\omega) + g (k-i\omega) & k^2 + \omega^2 + g^2 
    \end{pmatrix}.
\end{align}
Thus, we have 
\begin{widetext}
\begin{align}
    \Im(\tilde R(\omega)) & = \frac{\tilde R(\omega) - \tilde R^\dagger(\omega)}{2i}  = \frac{1}{|\Delta(\omega)|^2}\begin{pmatrix}
        \omega( k^2 + \omega^2 + g ) & (g+1) k\omega + \frac{i}{2} (g-1) (k^2 -\omega^2 - g)  \\
         (g+1) k\omega - \frac{i}{2} (g-1) (k^2 -\omega^2 - g)& \omega( k^2 + \omega^2 + g )
    \end{pmatrix},
\end{align}
\end{widetext}
which upon plugging in Eq.~\eqref{eq:FDT-violation} gives us the FDT violation matrix $\chi(g,\omega)$ as 
\begin{align}
    \chi(g,\omega) & = \frac{2D}{|\Delta(\omega)|^2}  \nonumber \\ & 
    ~~~\times \begin{pmatrix}
        1-g & -\frac{i(g-1)}{2\omega} (\omega^2 + k^2 -g) \\
        \frac{i(g-1)}{2\omega} (\omega^2 + k^2 -g) & g (g-1)
    \end{pmatrix}.
\end{align}
Clearly, $\chi(g=1,\omega)=0$, implying the expected equilibrium result.

Now, following Eq.~\eqref{eq:Individual-dissipated-heat}, we obtain the dissipated heat for individual particles to be 
\begin{align} \label{eq:heat-individual-Model-I}
    \langle Q_2\rangle & = \frac{Dg(g-1)}{2k},~
    \langle Q_1\rangle  =\frac{D(1-g)}{2k}.
\end{align}
Note that at $g=1$, we have $\langle Q_1\rangle= \langle Q_2\rangle = \langle Q_\text{tot}\rangle =0$, consistent with equilibrium. At $g=-1$, $\langle Q_1\rangle= \langle Q_2\rangle =D/k$. Moreover, note that at $g=0$, $\langle Q_1\rangle / \langle Q_2 \rangle $ diverges. This holds true even for larger $N$, as we will show later. 

\begin{figure*}
    \centering
    \includegraphics[width=1.\linewidth]{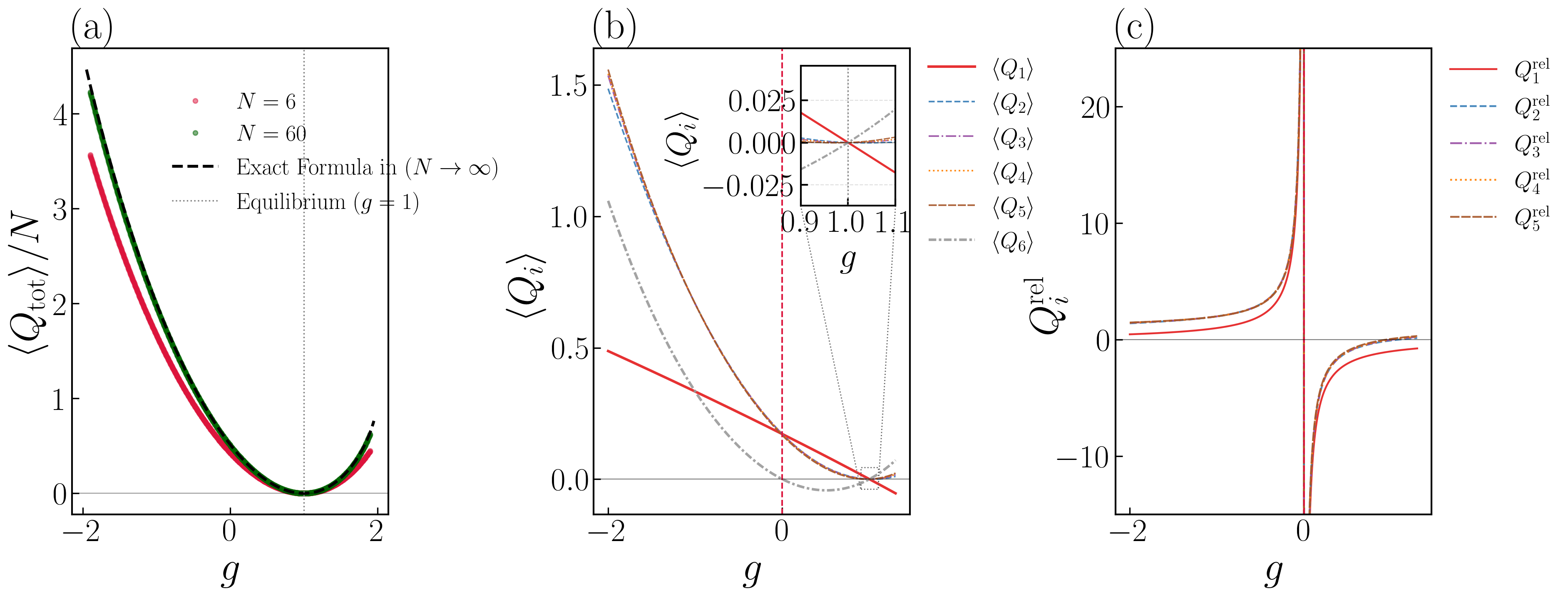}
    \caption{ Model~III: (a) Total heat dissipated per particle shown for two system sizes, compared against the thermodynamic expression in Eq.~\eqref{eq:Total-heat-thermodynamic-limit}.  (b) For system size $N=6$, the heat dissipation for individual particles is shown, with the inset showing the negative heat dissipation of boundary particles around $g=1$. (c) The relative heat dissipation of each particle with respect to the rightmost boundary is shown, with a clear signature of divergence at $g=0$. The parameters used in all panels are
    $k=3,~D=1$. The used data are based on exact results derived in Sec.~\ref{Sec:Heat-dissipation}.}
    \label{fig:Heat-dissipation}
\end{figure*}

\subsection{For Model II}
For a fixed realization of the drift matrix in Eq.~\eqref{eq:Drfit-matrix-model-II} of Model~II, focusing on the first dimer, namely, $(i,i+1)=(1,2)$, we obtain the response and the correlation matrix to be 
\begin{align}
    & \tilde R(\omega) = \frac{1}{\Delta(\omega)} \begin{pmatrix}
        k_2-i\omega & 1 \\
        g & k_1-i\omega
    \end{pmatrix}, \\
    & \tilde G(\omega)  = \frac{2D}{|\Delta(\omega)|^2 } \nonumber \\ & ~~~~\times \begin{pmatrix}
        k_2^2 + \omega^2 + g^2 & (k_1+i\omega) + g (k_2-i\omega) \\
        (k_1-i\omega) + g (k_2+i\omega) & k_1^2 + \omega^2 + 1
    \end{pmatrix}.
\end{align}
Plugging this back into Eq.~\eqref{eq:FDT-violation}, we obtain the FDT violation matrix as
\begin{align}
    \chi(g,\omega) & = \frac{2D}{|\Delta(\omega)|^2}  \nonumber \\ & 
    ~\times \begin{pmatrix}
         1-g &  -\frac{i(g-1)}{2\omega}  ( k_1k_2 -\omega^2  -g)  \\
         \frac{i(g-1)}{2\omega} ( k_1k_2 -\omega^2  -g) & g (g-1)
    \end{pmatrix}.
\end{align}
Putting together in Eq.~\eqref{eq:Individual-dissipated-heat}, we obtain the heat dissipation by the individual particles to be 
\begin{align}
     \langle Q_1\rangle = \frac{D(1-g)}{k_1+k_2},
    ~ \langle Q_2\rangle = \frac{Dg(g-1)}{k_1+k_2}.
\end{align}
Evidently, this is a simple generalization of Eq.~\eqref{eq:heat-individual-Model-I} for the case in which the individual constituents of the dimer have different stiffness constants. Indeed, one can show that for an arbitrary dimer, we have  
\begin{align}
     \langle Q_i\rangle = \frac{D(1-g)}{k_i+k_{i+1}},
    ~ \langle Q_{i+1}\rangle = \frac{Dg(g-1)}{k_i+k_{i+1}}.
\end{align}

\subsection{For Model III}
Next, we discuss the FDT violation for the drift matrix in Eq.~\eqref{eq:Model III-drift matrix} for Model~III. The correlation in Eq.~\eqref{eq:Correlation} and the response in Eq.~\eqref{eq:Response}, and thereby, the FDT violation matrix in Eq.~\eqref{eq:FDT-violation}, can be computed in a straightforward manner. Using Eq.~\eqref{eq:Individual-dissipated-heat}, let us focus on the dissipated heat for individual particles in Eq.~\eqref{eq:Individual-dissipated-heat}, which has a very simple form involving the drift matrix and the steady-state covariance matrix, as 
\begin{align}\label{eq:Simpler-form-dissipated-heat}
    \langle Q_i\rangle = (M C_{\text{ss}} M^T )_{ii} + DM_{ii}. 
\end{align}
A detailed derivation is outlined in Appendix.~\ref{APP:Derivation-off-heat-dissipation-shortcut}. Now the condition $\vec\nabla \cdot \vec J_{\text{ss}}(\vec x) =0$ gives us the following Lyapunov condition:
\begin{align}\label{eq:Lyapunov}
    M C_{\text{ss}} + C_{\text{ss}} M^T + 2D \mathbb{I} =0.
\end{align}
Denoting $A\equiv MC_{\text{ss}} $, we note that $A + A^T =-2D\mathbb{I}$ gives $A_{ii} = -D$ and $A_{ij} = -A_{ji}$. Using this property and Eq.~\eqref{eq:Model III-drift matrix}, we now obtain 
\begin{align}
    (M C_{\text{ss}} M^T)_{ii} & = \sum_{jl=1}^N M_{ij} (C_{\text{ss}})_{jl} M_{il} \nonumber \\
    & =\sum_{j=1}^N M_{ij} A_{ij} \nonumber \\
    & = -g A_{i-1,i} + kD + A_{i,i+1},
\end{align}
which leads to 
\begin{align}
    \langle Q_i\rangle =   A_{i,i+1} - g A_{i-1,i},
\end{align}
with $A_{i,i+1} = g (C_{\text{ss}})_{i-1,i+1} - k (C_{\text{ss}})_{i,i+1} + (C_{\text{ss}})_{i+1,i+1}$. 
Now, combining all these with $A_{N,N+1}=A_{0,1}=0$, the total dissipated heat $\langle Q_{\text{tot}} \rangle$ takes the following instructive form: 
\begin{align}
    \langle Q_{\text{tot}} \rangle & = \sum_{i=1}^N \langle Q_i \rangle = (1-g) \sum_{i=1}^{N-1} A_{i,i+1}, 
\end{align}
which immediately reveals that at equilibrium, corresponding to $g=1$, the dissipated heat to the bath is precisely zero. Moreover, in the limit $N \to \infty$, we find that the average heat dissipated per particle is 
\begin{align} \label{eq:Total-heat-thermodynamic-limit}
    \frac{\langle Q_{\text{tot} }\rangle}{N } = \frac{D(1-g)^2}{k + \sqrt{k^2 - (1+g)^2}}.
\end{align}
The derivation of this expression is detailed in Appendix~\ref{APP:Total heat-dissipation}. Clearly, as expected, the total heat dissipated at NESS into the bath is non-negative. In Fig.~\ref{fig:Heat-dissipation}(a), the approach to the thermodynamic expression in Eq.~\eqref{eq:Total-heat-thermodynamic-limit} is shown for two system sizes, and we get an excellent match for a system size as small as $N=60$.

Now, let us consider specifically the dissipated heat at the two boundaries: 
\begin{align} \label{eq:Boundary-heats}
    & \langle Q_1\rangle = (C_{\text{ss}})_{22} - k (C_{\text{ss}})_{12},\\
    & \langle Q_N \rangle = -g \left[ g (C_{\text{ss}})_{N-2,N} -k (C_{\text{ss}})_{N-1,N}  +  (C_{\text{ss}})_{N,N} \right].
\end{align}
This immediately reveals that $g=0$ yields $\langle Q_N \rangle =0$, implying that the relative heat $Q_i^{\text{rel}}$ dissipated with respect to the rightmost boundary,
\begin{align}
    Q_i^{\text{rel}} (g) \equiv \frac{\langle Q_i\rangle}{ \langle Q_N\rangle};~\forall\ i <N, 
\end{align}
diverges at EP, as shown in Fig.~\ref{fig:Heat-dissipation}(c). This particular feature has no equilibrium counterpart, and is attributed to the non-reciprocal nature of the underlying dynamics.

At  pseudo-equilibrium when $g=-1$, the steady-state covariance matrix is $C_{\text{ss}} = (D/k)\mathbb{I}$, leading to the heat-dissipation to be 
\begin{align} 
    \langle Q_1\rangle = \langle Q_N\rangle = \frac{D}{k}, \quad \langle Q_i \rangle_{1\le i \le N} = \frac{2D}{k}. 
\end{align}
As shown in Fig.~\ref{fig:Heat-dissipation}(b), the pseudo-equilibrium point therefore gives rise to a distinct spatial structure in the heat dissipation: the particles separate into two groups, with the boundary particles dissipating $D/k$ each, while the bulk particles dissipate $2D/k$ each. 

Moreover, the inset of Fig.~\ref{fig:Heat-dissipation}(b), together with Eq.~\eqref{eq:Boundary-heats}, reveals a remarkable distinction between the bulk and boundary particles. While all bulk particles dissipate heat into the bath at NESS, the boundary particles can, in contrast, absorb heat from their surroundings. Interestingly, the two boundary particles exchange their roles around $g=1$: the boundary that absorbs heat for $g<1$ dissipates heat for $g>1$, while the opposite occurs at the other boundary. This switch is clearly visible in the inset of Fig.~\ref{fig:Heat-dissipation}(b) for $N=6$. We have verified that this observation persists upon increasing the system size $N$. Such anomalous heat absorption at the boundary is a consequence of non-reciprocal interactions in our setup, and one may wonder whether this could potentially persist beyond the quadratic interactions considered here, and be more of a generic signature of non-reciprocity. For individual particles, the entropy production associated with this phenomenon can be readily inferred from the heat flux as $\langle Q_i\rangle/D$, which becomes negative at the boundary. Negative local entropy production of this kind has been discussed in the context of the minimal realization of Maxwell's demon~\cite{Loos2020}. Taken together, our observations suggest that boundary refrigeration may constitute a distinctive signature of non-reciprocity. Establishing the generality of this phenomenon, particularly for nonlinear interacting non-reciprocal systems, remains an interesting direction for future investigation.

\section{Conclusion and Outlook}
In summary, we have developed a hierarchy of non-reciprocally interacting Ornstein--Uhlenbeck processes and demonstrated that non-reciprocity provides a common framework for exceptional-point physics, anomalous relaxation, and nonequilibrium thermodynamics in classical stochastic systems. Across diverse models, we have identified the conditions under which the drift matrix assumes Jordan canonical forms and develops exceptional points of different orders, with the many-body model being directly related to the well-known Hatano--Nelson model. At an EP, the breakdown of diagonalizability replaces conventional exponential relaxation by polynomially dressed exponential dynamics, with the polynomial degree controlled by the order of the EP and, in extended systems, by system size. Remarkably, we have unveiled that the temporal response at an exceptional point can serve as a dynamical probe of spatial structure.

Our results further reveal that non-reciprocity can produce forms of nonequilibrium behaviour that are qualitatively distinct from conventional NESS. At the special value of the non-reciprocity parameter $g=-1$, the steady-state measure factorizes into single-particle distributions, furnishing an explicit example of a steady state that has the Gibbs form and yet is a genuine NESS. Moreover, the $N$-particle interacting OU dynamics gets mapped to $N/2$ non-interacting complex OU particles. Finally, by using the Harada--Sasa relation, we showed that the steady-state heat dissipation is strongly spatially structured. Moreover, even though the total heat dissipation remains positive, individual boundary particles can exhibit negative heat dissipation, allowing a boundary to function as an effective cold reservoir and thereby realizing boundary refrigeration, while the bulk consistently behaves as an effective hot reservoir. Taken together, these findings establish non-reciprocal OU processes as a simple and analytically tractable setting in which exceptional points, disorder-induced singularities, anomalous relaxation, pseudo-equilibrium, and spatially resolved nonequilibrium thermodynamics coexist. They suggest that the interplay between non-reciprocity and defective dynamics may provide a broader route toward controlling relaxation and energy transport in classical stochastic systems far from equilibrium.

This work opens up several directions for further investigation, both from a fundamental and an applied perspective.
\begin{itemize}

\item It is of immediate interest to investigate the robustness of the present results beyond the linear Gaussian setting. In particular, one may consider non-linear interaction forces, colored noise, and state-dependent noise amplitudes, and investigate about the robustness of the signatures identified here, such as the polynomial-in-time dressing of relaxation at exceptional points, the associated structure of cross-correlations, and the pseudo-equilibrium point. 

\item A natural direction for future work is to generalize the present framework from a one-dimensional chain to arbitrary networks, with each node hosting an Ornstein--Uhlenbeck process and interacting non-reciprocally with its connected neighbours. In this setting, the dynamics can be written as
\begin{align}
\dot{\vec{x}}=M\mathbf{x}+\vec{\eta},\qquad M=-K+A,
\end{align}
where $K=k \delta_{ij}$ contains the stiffness coefficients,  and $A$ is the weighted directed adjacency matrix encoding the inter-node non-reciprocal interactions. The underlying network topology then provides an additional control parameter, naturally raising the question of whether non-reciprocity can be harnessed to enhance targeted transport~\cite{Barabasi2016}.

\item The boundary refrigeration mechanism raises the possibility of exploiting non-reciprocity as a resource for stochastic heat engines and refrigerators~\cite{Martinez2017}. It is natural to ask whether it can enhance refrigeration performance, enable cooling where reciprocal systems fail, or improve the efficiency of otherwise inefficient stochastic machines, while characterizing the trade-offs with entropy production and coupling strength.

\end{itemize}

\begin{acknowledgments}
The authors acknowledge Rupak Majumder for discussions and collaboration on related topics. This work is supported by the Department of Atomic Energy, Government of India, under Project Identification Number RTI-4012 and RTI-4013. Computations were carried out on the computing clusters at the Department of Theoretical Physics, TIFR, Mumbai. We thank Ajay Salve and Kapil Ghadiali for their computational support.
\end{acknowledgments}

\appendix
\section{Hermitian matrices cannot host exceptional points (EP's)} \label{APP:EP-absence-Hermitian}
Considering any matrix $A \in \mathbb{C}^{(n\times n)} $ with $\mathbb{C}^n$ being the complex vector space, for any eigenvalue $\lambda = E$, one can define a Kernel vector space for $B = A - E \mathbb{I}$ as 
\begin{align}
    Ker(B) \equiv \{ \forall |v \rangle \in Ker(B) : B |v \rangle = 0 \}.
\end{align}
Clearly, $Ker (B) \subseteq \mathbb{C}^{n} $. 
Now, if $\lambda =E$ is $m$-fold degenerate, then the charateristic polynomial for $A$ must satisfy 
\begin{align}
    p(\lambda) = (\lambda - E)^m f (\lambda); ~ f(\lambda = E) \neq 0. 
\end{align}
Henceforward, $m$ will be referred to as the algebraic multiplicity for $\lambda = E$. Now, let us call the dimension of $Ker(B)$ the geometric multiplicity, and denote it by $d$. 
Evidently, we have  
\begin{align}
    d \le m. 
\end{align}

The condition for the existence of EP's is therefore pinned by the strict inequality 
\begin{align}
    d < m,
\end{align}
which essentially signifies that the eigenvectors have coalesced, and one loses some directions in the vector space. 
To account for the lost eigenvectors, we extend the notion of $Ker(B)$ to a general one
\begin{align}
    Ker(B^k)  \equiv \{ \forall |v \rangle \in Ker(B^k) : B^k |v \rangle = 0 \},
\end{align}
with the implication that there may exist some vector $|u \rangle$ such that $B^k|u \rangle=0$ but $B|u \rangle \ne 0$. This automatically gives us 
\begin{align}
    Ker(B^k) \subseteq Ker(B^{k-1}) \subseteq \ldots \subseteq Ker(B).
\end{align}
Now, to this end, we introduce the notion of generalized eigenvectors ($B^k|u \rangle=0$; $B|u \rangle \ne 0$ ), which will serve as the proxy for the lost eigenvectors in $Ker(B)$. The systematic way to find them is the following iterative process: 
\begin{itemize}
    \item If $m > d$, then there are $d$ linearly independent eigenvectors in $Ker(B)$. Lets denote them by $\{ | v_i \rangle \}_{1\le i \le d}$.
    \item Now, one can perform the following 
    \begin{align}
        B | u_i \rangle = |v_i\rangle,
    \end{align}
    where the independent $\{|u_i\rangle\}$s are the generalized eigenvectors. Let us denote the dimension of $Ker(B^2)$ to be $d_2$. 
    \item For $m>d$, it is guranteed that $d_2>d$. This follows from the stabilization theorem, which we will discuss below. 

    \item If $d_2<m$, we iterate the process $k$ times, i.e., 
    \begin{align}
        B | \tilde{u}_i \rangle = |\tilde{v}_i \rangle; \{|\tilde{v}_i\}_{1\le i \le d_{k-1}} \rangle \in Span( Ker(B^{k-1}) ), 
    \end{align}
     such that the dimension of $Ker(B^k)$, $d_k =m$. 
\end{itemize}
This will generate the full set of eigenvectors. One is guaranteed to find all the eigenvectors since one can show that for the vector space 
\begin{align}
    \mathcal{G} = \cup_{k} Ker(B^k),
\end{align}
its dimension $Dim ( \mathcal{G}) = m$. Additionally, if one has to wait till $Ker(B^k)$ to fully recover the algebraic multiplicity $m$, then 
\begin{align}
    d_k > d_{k-1} > \ldots > d_2 > d. 
\end{align}
This follows directly from the following theorem.\\

\textit{\textbf{Stabilization theorem:---}} If $Ker(B^{k+1}) =  Ker(B^{k})$, then $Ker(B^{k+r}) =  Ker(B^{k})$.\\

\underline{\textit{Proof:---}} 
We know $Ker(B^{k+1}) \subseteq Ker(B^{k+2})$. 
Clearly, 
\begin{align}
    & \forall | v \rangle \in Ker(B^{k+2}) \\
    & \implies B^{k+2} | v \rangle =0 \\
    & \implies B^{k+1} (B| v \rangle) =0\\
    & \implies B| v \rangle \in Ker(B^{k+1}).
\end{align}
Since $Ker(B^{k+1}) = Ker(B^{k})$, one has $B| v \rangle  \in Ker(B^{k})$. In turn, this implies that 
\begin{align}
    & B^{k+1} | v \rangle =0 \\
    & \implies | v \rangle \in Ker(B^{k+1}).
\end{align}
So, $\forall | v \rangle \in Ker(B^{k+2})$, it is therefore true that $| v \rangle \in Ker(B^{k+1})$. Hence, 
\begin{align}
    Ker(B^{k+2}) = Ker(B^{k+1}) = Ker(B^{k}).
\end{align}
Therefore, trivially, it leads to 
\begin{align}
     Ker(B^{k+r}) = \ldots = Ker(B^{k+2}) = Ker(B^{k+1}) = Ker(B^{k}). 
\end{align}

\subsubsection{Why Hermitian matrices cannot have EP's?}
From the stabilization theorem above, it is therfore suffices to show that for Hermitian systems, $Ker(B^2) \neq Ker(B)$ to signal the absence of EP's.

\textit{\textbf{Claim:---}} For Hermitian systems, $Ker(B^2) = Ker(B)$.\\

\underline{\textit{Proof:---}} We use the method of contradiction. 

Let us assume that there exists a $| v \rangle \in Ker (B^2)$ such that $B| v \rangle \ne 0 $. We note that $B^\dagger = B$.
Now since $B^2 | v \rangle =0 0$, we have 
\begin{align}
    & \langle v | B^2 | v\rangle =0 \\
    & \implies \langle v | B^\dagger B  | v\rangle =0 \\
    & \implies \langle (Bv) | (Bv)\rangle =0 \\
    & \implies || B|v\rangle|| =0.
\end{align}
This can be true if and only if $B |v\rangle =0$, which clearly contradicts the assumption that $B |v\rangle \ne 0 $. 
Therefore, for Hermitian systems, $Ker(B^2) = Ker (B)$. 

\section{Spectrum of Hatano-Nelson model in Eq.~\eqref{eq:Hatano-Nelson}} \label{APP:Hatano-Nelson-Spectrum}

For the Hamiltonian $\hat H_{\text{HN}}$ in Eq.~\eqref{eq:Hatano-Nelson}, let us assume that the eigenstate $|\Psi\rangle=\sum_{j=1}^N \psi_j |j \rangle$ corresponds to the eigenvalue $E$ such that $\hat H_{\text{HN}} | \Psi\rangle = E |\Psi\rangle $, leading to 
\begin{align} \label{eq:App-eigen-eq}
    \psi_{j+1} + g \psi_{j-1} - k \psi_j =E \psi_j ,
\end{align}
with the boundary conditions, $\psi_0 =0 $ and $\psi_{N+1}=0$.

Let us now introduce a gauge transformation of the following 
\begin{align}
    \psi_j \rightarrow g^{j/2} \phi_j
\end{align}
such that Eq.~\eqref{eq:App-eigen-eq} leads to
\begin{align}
    \phi_{j+1} + \phi_{j-1} = \frac{(E +k )}{\sqrt{g}} \phi_j,
\end{align}
with the boundary conditions $\phi_0 =0$ and $\phi_{N+1}=0$. This is a simple second-order difference equation, and one can take $\phi_j = e^{iqj} (\text{or } e^{-iqj})$ as a trial solution, leading to 
\begin{align}
    E= -k + 2\sqrt{g} \cos{q}.
\end{align}
A generic linear-superposition solution would be $\phi_j = A \cos{(qj)} + B \sin{(qj)}$; the boundary conditions will determine the coefficients and quantizations. Since $\phi_0 =0$, we get $A=0$, and $\phi_{N+1}=0$ gives us 
\begin{align}
    q_n = \frac{n\pi}{N+1}, \forall n=1,\ldots, N.
\end{align}
$B$ can be fixed from the normalization condition. Therefore, the spectrum of the Hatano-Nelson model is simply 
\begin{align} \label{eq:App-spectrum-Hatano-Nelson}
    E_n = -k + 2\sqrt{g} \cos{\biggr( \frac{n\pi}{N+1}\biggr)}.
\end{align}

\section{Derivation of Eq.~\eqref{eq:Exponential-matrix-elements}} \label{APP:Derivation-exponential-Hatano-Nelson}

We derive the matrix elements of the propagator $e^{M(g)t}$ appearing in
Eq.~\eqref{eq:Exponential-matrix-elements}. Starting from
\begin{align}
M(g)=-k\mathbb{I}_N+T(g),
\end{align}
and using the similarity transformation
\begin{align}
T(g)=\mathcal{D}(\sqrt{g}\,A)\mathcal{D}^{-1},
\end{align}
we can write
\begin{align}
M(g)
=
\mathcal{D}
\left(-k\mathbb{I}_N+\sqrt{g}\,A\right)
\mathcal{D}^{-1}.
\end{align}
Since the matrix exponential is invariant under similarity transformations,
we have
\begin{align}
e^{M(g)t}
=
\mathcal{D}
e^{(-k\mathbb{I}_N+\sqrt{g}\,A)t}
\mathcal{D}^{-1}.
\end{align}
Furthermore, since $\mathbb{I}_N$ commutes with $A$, this reduces to
\begin{align}
e^{M(g)t}
=
e^{-kt}\mathcal{D}
e^{\sqrt{g}At}
\mathcal{D}^{-1}.
\label{eq:app_propagator_similarity}
\end{align}

It therefore remains to evaluate the exponential of the symmetric
tridiagonal matrix
\begin{align}
A=
\sum_{l=1}^{N-1}
\left(
|l\rangle\langle l+1|
+
|l+1\rangle\langle l|
\right).
\end{align}
For open boundary conditions, the normalized eigenvectors of $A$ are
\begin{align}
|m\rangle
=
\sqrt{\frac{2}{N+1}}
\sum_{r=1}^{N}
\sin(r\theta_m)|r\rangle,
\qquad
\theta_m=\frac{m\pi}{N+1},
\end{align}
with corresponding eigenvalues
\begin{align}
\lambda_m=2\cos(\theta_m),
\qquad
m=1,\ldots,N.
\end{align}
Indeed, using
\begin{align}
\sin[(r-1)\theta_m]
+
\sin[(r+1)\theta_m]
=
2\cos(\theta_m)\sin(r\theta_m),
\end{align}
one readily verifies that
$A|m\rangle=\lambda_m|m\rangle$.

The spectral decomposition of $A$ consequently gives
\begin{align}
A
=
\sum_{m=1}^{N}
2\cos(\theta_m)
|m\rangle\langle m|,
\end{align}
and hence
\begin{align}
e^{\sqrt{g}At}
=
\sum_{m=1}^{N}
e^{2\sqrt{g}t\cos(\theta_m)}
|m\rangle\langle m|.
\label{eq:app_spectral_exp}
\end{align}
Taking the matrix element between the site states $|i\rangle$ and
$|j\rangle$, and using
\begin{align}
\langle i|m\rangle
=
\sqrt{\frac{2}{N+1}}\sin(i\theta_m),
\end{align}
we obtain
\begin{align}
\left(e^{\sqrt{g}At}\right)_{ij}
&=
\langle i|e^{\sqrt{g}At}|j\rangle
\nonumber\\
&=
\sum_{m=1}^{N}
e^{2\sqrt{g}t\cos(\theta_m)}
\langle i|m\rangle
\langle m|j\rangle
\nonumber\\
&=
\frac{2}{N+1}
\sum_{m=1}^{N}
e^{2\sqrt{g}t\cos(\theta_m)}
\sin(i\theta_m)\sin(j\theta_m).
\label{eq:app_A_exponential}
\end{align}

Finally, since $\mathcal{D}$ is diagonal,
\begin{align}
(\mathcal{D}^{-1})_{jj}=g^{-(j-1)/2},
\end{align}
and Eq.~\eqref{eq:app_propagator_similarity} gives
\begin{align}
\left(e^{M(g)t}\right)_{ij}
&=
e^{-kt}
\mathcal{D}_{ii}
\left(e^{\sqrt{g}At}\right)_{ij}
(\mathcal{D}^{-1})_{jj}
\nonumber\\
&=
e^{-kt}
g^{(i-1)/2}g^{-(j-1)/2}
\left(e^{\sqrt{g}At}\right)_{ij}
\nonumber\\
&=
e^{-kt}g^{(i-j)/2}
\left(e^{\sqrt{g}At}\right)_{ij}.
\end{align}
Substituting Eq.~\eqref{eq:app_A_exponential}, we finally obtain
\begin{align}
\left(e^{M(g)t} \right)_{ij}
& =
e^{-kt}g^{(i-j)/2}
\frac{2}{N+1} \nonumber \\
& \sum_{m=1}^N
e^{2\sqrt{g}t\cos(\theta_m)}
\sin(i\theta_m)\sin(j\theta_m)
\label{eq:app_final_propagator}
\end{align}
where
\begin{align}
\theta_m=\frac{m\pi}{N+1}.
\end{align}
This completes the derivation of Eq.~\eqref{eq:Exponential-matrix-elements}.

\section{Fluctuation-dissipation theorem (FDT) in higher dimension} \label{APP:FDT derivation}
Let us outline the standard set-up for the linear-response theory. Let us assume that $\vec{h}(t)$ is a time-dependent additive perturbation in the system such that the dynamics reads as
\begin{align} \label{eq:perturbed OU}
    \dot{\vec{x}} \equiv M \vec{x} + \vec{\eta}(t) + \vec{h}(t),
\end{align}
where $M$ is the drift matrix and the noise has the correlator $\langle \vec\eta(t) \vec\eta^T(t')  \rangle = 2D\delta(t-t') \mathbb{I} $. 

Let us define the response matrix to the perturbation $\vec{h}(t) \to 0$ as 
\begin{align}
    R_{ij} (t,t') = \frac{\delta \langle x_i (t)\rangle }{\delta  h_j(t')   }\biggr|_{\vec{h}=0}, 
\end{align}
such that 
\begin{align}
     \langle \vec{x}(t)\rangle_{\vec{h}} - \langle \vec{x}(t)\rangle_{\vec{h}=0} = \int_{-\infty}^t dt' R(t,t') ~\vec{h}(t') + \mathcal{O}(h^2).   
\end{align}

Now, returning to Eq.~\eqref{eq:perturbed OU}, we first perform the noise average and subsequently take the functional derivative with respect to the perturbation field, yielding the evolution equation for the response matrix as
\begin{align}
      \frac{d}{dt} \left[ R_{ij}(t,t')  \right] = \sum_{k} M_{ik} R_{kj}(t,t') + \delta_{ij} \delta(t-t'). 
\end{align}

Now let us integrate in time from $t' -\epsilon $ to $t' + \epsilon $ with $\epsilon\to 0 $. This yields
\begin{align}
    & R_{ij} (t'+\epsilon,t') - \underbrace{R_{ij}(t'-\epsilon,t')}_{=0\text{~due to causality}} \nonumber \\ 
    & =  \sum_{k} M_{ik} \underbrace{\int_{t'-\epsilon}^{t' + \epsilon} dt~  R_{kj}(t,t')}_{=0 \text{ in the limit $\epsilon \to 0$}} +  \delta_{ij};
\end{align}
implying
\begin{align}
   R_{ij} (t',t') = \delta_{ij}. 
\end{align}
Equipped with this initial condition, the linear evolution equation can be solved easily with the causality included, yielding the solution 
\begin{align}
    R(t,t') = \Theta(t-t') e^{M(t-t')} \mathbb{I} = R(t-t'). 
\end{align}
Reverting back to the evolution, we can now write it as 
\begin{align}
   &  \frac{d}{dt} R(t-t') = M R(t-t') + \mathbb{I} \delta(t-t').
   \end{align}
Defining $\tau=t-t'$, we have
\begin{align}
\frac{d}{d\tau} R(\tau) = M R(\tau) + \mathbb{I} \delta(\tau), 
\end{align}
which, upon Fourier transformation to the frequency space, yields
\begin{align}
\tilde{R}(\omega) = [-i\omega - M]^{-1}. 
\end{align}
This completes the derivation of the response matrix in Eq.~\eqref{eq:Response} of the main text.

Now, the two-point correlation matrix in time is defined as 
\begin{align}
    G(t,t') \equiv \langle \vec{x}(t) \vec{x}^T(t') \rangle,  
\end{align}
which, upon using $\vec{x}(t) = \int_{-\infty}^{t} dt' ~R(t,t') \vec{\eta}(t')$, yields
\begin{align}
    & G(t,t') = \int_{-\infty}^{t} ds\int_{-\infty}^{t'} ds' R(t,s) \langle \vec{\eta}(s) \vec{\eta}^T(s') \rangle R^T(t',s'). 
\end{align}  
At NESS, due to time translation invariance, we obtain
\begin{align}
     G(\tau) = 2D \int_{-\infty}^{\tau} ds\int_{-\infty}^{\infty} ds' R(\tau-s) \delta(s-s') R^T (-s').
\end{align}
Now, in the frequency space via Fourier transformation, we finally get
\begin{align}     
         \tilde{G}(\omega) & = 2D \tilde{R}(\omega) \tilde{R}^\dagger(\omega) \nonumber 
      & = 2D [-i\omega - M]^{-1} [i\omega - M^T]^{-1}.
\end{align}

The above expression is used in Eq.~\eqref{eq:Correlation} of the main text, and it captures the relation of correlation and response functions in the regime of linear response. Now, the fluctuation-dissipation theorem tells us that at equilibrium, they must satisfy the following relation  
\begin{align}
    \tilde G(\omega) = \frac{2D}{\omega} \Im(\tilde R(\omega)). 
\end{align}
Away from equilibrium, this equality breaks down. To quantify this breakdown, we introduce the FDT violation matrix as 
\begin{align}
    \chi(\omega) = \tilde G(\omega) - \frac{2D}{\omega} \Im(\tilde R(\omega)),
\end{align}
which has been used in Eq.~\eqref{eq:FDT-violation} of the main text.
\section{Derivation of Eq.~\eqref{eq:Simpler-form-dissipated-heat}} \label{APP:Derivation-off-heat-dissipation-shortcut}
From Eq.~\eqref{eq:Response} in the main text, we have the following 
\begin{align}
    M \tilde{R}(\omega) & = -i\omega \tilde{R}(\omega)  - \mathbb{I},
\end{align}
leading to 
\begin{align}
    (M \tilde{R}(\omega)) (M \tilde{R}(\omega))^\dagger & = \omega^2 \tilde{R}(\omega) \tilde{R}(\omega)^\dagger + i\omega \tilde{R}(\omega) \nonumber \\
    & ~~~~~~~~~- i\omega \tilde{R}(\omega)^\dagger + \mathbb{I} \nonumber \\
    & =\omega^2 \tilde{R}(\omega) \tilde{R}(\omega)^\dagger -2 \omega \Im[\tilde{R}(\omega)] + \mathbb{I}. 
\end{align}
Now, recall that the stationary covariance matrix $C_{\text{ss}}$ is simply 
\begin{align}
    C_{\text{ss}} (\omega) = 2D \int \frac{d\omega}{2\pi } \tilde{R}(\omega) \tilde{R}^\dagger (\omega). 
\end{align}
Therefore, $\int \frac{d\omega}{2\pi } (M \tilde{R}(\omega)) (M \tilde{R}(\omega))^\dagger = (M C_{\text{ss}}M^T)/(2D)$, and we obtain 
\begin{align}
    ( M C_{\text{ss}}M^T)_{ii} & = 2D \int \frac{d\omega}{2\pi } \nonumber \\
    & \times \left[ \omega^2 (\tilde{R}(\omega) \tilde{R}(\omega)^\dagger)_{ii} -2 \omega(\Im[\tilde{R}(\omega)])_{ii} + 1 \right].
\end{align}

Now subtracting $( M C_{\text{ss}}M^T)_{ii}$ from $\langle Q_i\rangle$, we obtain 
\begin{align}
    \langle Q_i\rangle - ( M C_{\text{ss}}M^T)_{ii} = 2D \int \frac{d\omega}{2\pi } \left[\omega (\Im[\tilde{R}(\omega)])_{ii} -1  \right]
\end{align}
To evaluate
\begin{align}
\int_{-\infty}^{\infty} \frac{d\omega}{2\pi}
\left[
\omega\,\Im(\tilde{R})_{ii}(\omega)-1
\right],
\end{align}
we first note that
\begin{align}
\Im(\tilde{R}_{ii})(\omega)
=
\frac{1}{2i}
\left[
\tilde{R}_{ii}(\omega)-\tilde{R}_{ii}(-\omega)
\right].
\end{align}
This allows us to rewrite the integrand as
\begin{align}
\frac{1}{2}
\left[
-i\omega \tilde{R}_{ii}(\omega)-1
+
i\omega \tilde{R}_{ii}(-\omega)-1
\right].
\end{align}
Changing variables $\omega\to-\omega$ in the second term shows that the integral reduces exactly to
\begin{align}
\int_{-\infty}^{\infty}\frac{d\omega}{2\pi}
\left[
-i\omega \tilde{R}_{ii}(\omega)-1
\right].
\end{align}

Since
\begin{align}
\tilde{R}(z)
=
(-izI-M)^{-1}
=
\frac{i}{z}I
-\frac{1}{z^2}M
+\mathcal{O}(z^{-3})
\end{align}
for large $|z|$, the diagonal component behaves as
\begin{align}
-iz\tilde{R}_{ii}(z)-1
=
\frac{iM_{ii}}{z}
+\mathcal{O}(z^{-2}).
\end{align}

Because all eigenvalues $\lambda_\alpha$ of the stable drift matrix $M$ satisfy
\begin{align}
\mathrm{Re}(\lambda_\alpha)<0,
\end{align}
all poles
\begin{align}
z_\alpha=-i\lambda_\alpha
\end{align}
of $\tilde{R}(z)$ lie strictly in the upper complex half-plane,
\begin{align}
\mathrm{Im}(z_\alpha)>0.
\end{align}
We can therefore close the integration contour along a large semicircle $C_R$ in the pole-free lower half-plane, $\mathrm{Im}(z)<0$, oriented clockwise. By Cauchy's Residue Theorem, the real-line integral plus the arc integral vanishes.

On the large semicircle, only the leading asymptotic term contributes:
\begin{align}
\int_{C_R}\frac{dz}{2\pi}
\frac{iM_{ii}}{z}
=
\frac{iM_{ii}}{2\pi}(-i\pi)
=
\frac{1}{2}M_{ii}.
\end{align}
Hence,
\begin{align}
\int_{-\infty}^{\infty}
\frac{d\omega}{2\pi}
\left[
\omega\,\mathrm{Im}\,\tilde{R}_{ii}(\omega)-1
\right]
=
\frac{1}{2}M_{ii}.
\end{align}
Combining all this, we finally have 
\begin{align}
    \langle Q_i\rangle = (M C_{\text{ss}}M^T)_{ii} + D M_{ii}.  
\end{align}

\section{Derivation of Eq.~\eqref{eq:Total-heat-thermodynamic-limit}} \label{APP:Total heat-dissipation}

From Eq.~\eqref{eq:Simpler-form-dissipated-heat} in the main text, we have 
\begin{align}
    \langle Q_i\rangle = (M C_{\text{ss}} M^T )_{ii} + DM_{ii},
\end{align}
which, upon using Eq.~\eqref{eq:Lyapunov}, yields 
\begin{align} \label{eqAPP:Q_tot}
    \langle Q_\text{tot}\rangle = \sum_{i=1}^N \langle Q_i\rangle = N k D - Tr\left[ M^2 C_{\text{ss}} \right].
\end{align}
Now, let us first consider that for the Model~III, we have PBC instead of OBC, modifying to the drift matrix in Eq.~\eqref{eq:Model III-drift matrix} with the elements $M_{1,N}=1$ and $M_{N,1}=g$. Evidently, in the thermodynamic limit $N\to \infty$, the boundary conditions will not affect $\langle Q_{\text{tot}}\rangle$, which will yield the same dependence on $g$. The PBC assumption allows a direct transformation in the Fourier space, since the drift matrix can be written as 
\begin{align}
    M = \sum_{j=1}^N \left[ -k |j \rangle \langle j | + |j \rangle \langle j+1| + g |j+1\rangle \langle j| \right],
\end{align}
where $|j\rangle $ is the position-space vector, which, upon Fourier transformation defined by the matrix $\mathcal{F}$, yields
\begin{align}
    |j\rangle = \frac{1}{\sqrt{N}} \sum_{q} e^{-iqj} |q \rangle, 
\end{align}
with the inverse 
\begin{align}
    |q\rangle = \frac{1}{\sqrt{N}} \sum_{j=1}^N e^{iqj} |j \rangle. 
\end{align}
In the Fourier basis, the drift matrix becomes diagonal, i.e., $\mathcal{F} M \mathcal{F}^{-1}= \Lambda$, where $\Lambda = \text{Diag}\{\lambda(q) \}$ with 
\begin{align}
    M |q\rangle = ( -k + e^{iq} + ge^{-iq}) |q\rangle,
\end{align}
and $\lambda(q) = -k + (1+g) \cos{q} +i(1-g) \sin{q} \equiv -a(q) + i b(q) $, where $a(q) = k-(1+g) \cos{q}$, $b(q) =(1-g)\sin{q} $. 

Next, we obtain the steady-state covariance matrix in the Fourier space $C_{\text{ss}}^F$ following the Lyapunov equation in Eq.~\eqref{eq:Lyapunov} to be 
\begin{align}
    & \Lambda C_{\text{ss}}^F + C_{\text{ss}}^F \Lambda^\dagger + 2D \mathbb{I}_N=0, \nonumber \\
    \implies & C_{\text{ss}}^F(q,q') = \frac{-2D \delta_{q,q'}}{\lambda(q) + \lambda^*(q')}, 
\end{align}
implying that the covariance matrix in the Fourier space is also diagonal. 

Plugging this back into Eq.~\eqref{eqAPP:Q_tot} and dividing by $N$, we have the followig expression of the total heat dissipated per particle: 
\begin{align}
    \frac{\langle Q_\text{tot}\rangle}{N} = kD - \frac{1}{N} \sum_{q} \lambda^2(q) C_{\text{ss}}^F(q,q),
\end{align}
with $C_{\text{ss}}^F(q,q)= D/a(q)$. Now, in the thermodynamic limit $N\to \infty$, we have 
\begin{align}
    \frac{1}{N} \sum_{q} \longrightarrow \frac{1}{2\pi} \int_{-\pi}^\pi \mathrm{d}q, 
\end{align}
leading to 
\begin{align}
    \frac{\langle Q_\text{tot}\rangle}{N} = kD - \frac{D}{2\pi} \int_{-\pi}^\pi \mathrm{d}q~\biggr[ a(q) - \frac{b^2(q)}{a(q)} - 2i b(q)\biggr].
\end{align}
Now, note that $(1/2\pi)\int_{-\pi}^\pi \mathrm{d}q~ b(q) =0$, and $(1/2\pi)\int_{-\pi}^\pi \mathrm{d}q~ a(q)= k$. Combining these, we get 
\begin{align}
    \frac{\langle Q_\text{tot}\rangle}{N} & = \frac{D(1-g)^2}{2\pi} \int_{-\pi}^\pi \mathrm{d}q~ \frac{\sin^2{q}}{k - (1+g)\cos{q}} \nonumber \\
    & = \frac{D(1-g)^2}{k + \sqrt{k^2 -(1+g)^2}},
\end{align}
which is the expression in Eq.~\eqref{eq:Total-heat-thermodynamic-limit} in the main text.

\bibliography{Main}

\end{document}